\documentclass[journal,twoside]{IEEEtran}
\usepackage{amssymb}
\usepackage{graphicx}
\usepackage{bm}
\usepackage{algorithm}
\usepackage{algorithmic}
\usepackage{tabularx}
\usepackage{array}
\usepackage{booktabs}
\usepackage{soul}
\usepackage{epstopdf}
\usepackage{stfloats}
\usepackage{booktabs}  
\usepackage{cite}

\ifCLASSINFOpdf
\else
\fi
\usepackage[cmex10]{amsmath}
\usepackage{array}

\usepackage{mdwmath}
\usepackage{mdwtab}

\usepackage[tight,footnotesize]{subfigure}
\usepackage{color}
\makeatletter
\renewcommand{\maketag@@@}[1]{\hbox{\m@th\small\normalfont#1}}
\makeatother

\begin{document}

\title{Movable Antenna-Assisted Integrated Sensing and Communication Systems}
%

%


\author{\begin{normalsize}Chengjun~Jiang,~\IEEEmembership{\begin{normalsize}Graduate Student Member,~IEEE\end{normalsize}}, 
		Chensi~Zhang,~\IEEEmembership{\begin{normalsize}Member,~IEEE\end{normalsize}},
		Chongwen~Huang,~\IEEEmembership{\begin{normalsize}Member,~IEEE\end{normalsize}},
		Jianhua~Ge,~
		Dusit~Niyato,~\IEEEmembership{\begin{normalsize}Fellow,~IEEE\end{normalsize}},
		and~Chau~Yuen,~\IEEEmembership{\begin{normalsize}Fellow,~IEEE\end{normalsize}}\end{normalsize}

\thanks{Chengjun Jiang, Chensi Zhang, and Jianhua Ge are with the State Key Lab. of Integrated Service Networks, Xidian University, Xi'an, China (e-mail: cjjiang@stu.xidian.edu.cn, cszhang@xidian.edu.cn, jhge@xidian.edu.cn).
	
Chongwen Huang is with College of Information Science and Electronic Engineering, Zhejiang University, with the State Key Laboratory of Integrated Service Networks, Xidian University, and Zhejiang Provincial Key Laboratory of Info. Proc., Commun. \& Netw. (IPCAN), China. (e-mail: chongwenhuang@zju.edu.cn).

Dusit Niyato is with the College of Computing and Data Science, Nanyang Technological University, Singapore (e-mail: dniyato@ntu.edu.sg)

Chau Yuen is with the School of Electrical and Electronics Engineering, Nanyang Technological University, Singapore (e-mail: chau.yuen@ntu.edu.sg).

} 

}
\maketitle

\begin{abstract}
Movable antennas (MAs) enhance flexibility in beamforming gain and interference suppression by adjusting position within certain areas of the transceivers. In this paper, we propose an MA-assisted integrated sensing and communication framework, wherein MAs are deployed for reconfiguring the channel array responses at both the receiver and transmitter of a base station. Then, we develop an optimization framework aimed at maximizing the sensing signal-to-interference-plus-noise-ratio (SINR) by jointly optimizing the receive beamforming vector, the transmit beamforming matrix, and the positions of MAs while meeting the minimum SINR requirement for each user. To address this nonconvex problem involving complex coupled variables, we devise an alternating optimization-based algorithm that incorporates techniques including the Charnes-Cooper transform, second-order Taylor expansion, and successive convex approximation (SCA). Specifically, the closed form of the received vector and the optimal transmit matrix can be first obtained in each iteration. Subsequently, the solutions for the positions of the transmit and receive MAs are obtained using the SCA method based on the second-order Taylor expansion. The simulation results show that the proposed scheme has significant advantages over the other baseline schemes. In particular, the proposed scheme has the ability to match the performance of the fixed position antenna scheme while utilizing fewer resources.
\end{abstract}

\begin{IEEEkeywords}
Integrated sensing and communication, movable antenna, antenna position optimization, beamforming.
\end{IEEEkeywords}

%
\IEEEpeerreviewmaketitle

\section{Introduction}

\IEEEPARstart{T}{he blueprint} for sixth-generation (6G) wireless technology is increasingly becoming defined, driven by the pressing demands of the future industry landscape. For instance, advancements such as smart transportation and autonomous driving, the industrial internet of things, smart manufacturing, augmented reality, and virtual reality necessitate not only high-quality wireless connectivity but also highly accurate and robust sensing capabilities. As a result, in addition to wireless communications, a prevailing theme among visionary projections for 6G networks is the enhanced role of sensing, anticipated to be more critical than in any previous generation\cite{r11}. Further heightening the anticipation, future radio sensing and communication systems are converging in terms of hardware architectures, channel characteristics, and signal processing techniques, offering a unique opportunity to merge these capabilities in networks that transcend traditional communications paradigms\cite{r12}. In light of these developments, integrated sensing and communications (ISAC), which serves as a unified platform for fully integrating radar sensing and communication capabilities, has garnered significant attention from both academia and industry\cite{r13,r14,r15,r16,r29,r30}.

The realization of ISAC has been facilitated by significant advancements in technologies such as massive multiple input multiple output (mMIMO)\cite{r17} and millimeter wave (mmWave)\cite{r18}. The mMIMO antenna arrays can be significantly miniaturized owing to the reduced signal wavelengths, while mmWave signals benefit from enhanced transmission distances enabled by the beamforming gains of these arrays. Consequently, the employment of large bandwidths and antenna arrays not only enhances communication capacity and numerous connections but also significantly improves radar capabilities in terms of range and angular resolution. In other words, the degrees of freedoms (DoFs) provided by MIMO technology are crucial for achieving the dual functionality of high-performance communication and precise sensing in ISAC systems\cite{r19}. In \cite{r20}, Liu \emph{et} al. proposed a joint transmit beamforming model for a dual-functional MIMO radar and multiuser MIMO communication transmitter, which achieved sensing and communication by constructing multiple beams directed at the radar target and the communication receiver. Obviously, in order to make a dual-function radar communication (DFRC) base station (BS) perform well in ISAC, the DoFs of the system are crucial, which makes the need for antenna array size more urgent.

However, the large antenna dimensions in mMIMO systems present significant challenges in terms of power consumption and radio resource management in practical applications. To this end, in addition to optimizing hardware design, improving algorithms, and intelligent resource management, the introduction of new technologies is also imperative and feasible. For example, reconfigurable intelligent surfaces (RIS) have been demonstrated to be effective for ISAC systems\cite{r36,r21,r22,r50,r51}, which can reconfigure the transmission path. RIS can reduce the number of antennas and radio-frequency (RF) links while maintaining the performance. Recently, the moveable antenna (MA) technique has received significant academic interest due to its capability to dynamically adjust its position, orientation, and configuration\cite{r3,r2,r31,r32,r33,r34,r35,r38}. This adaptability enables the antenna to reconfigure transmission channels by altering its physical position and orientation, thus optimizing signal reception and transmission.

MA is advocated to overcome the limitations of conventional MIMO systems, which are unable to fully exploit the spatial variations of wireless channels in a specified transmit and receive field due to fixed antenna positions. Specifically, the MA is interfaced with the RF chain through a flexible cable, allowing it to move within the spatial area via a driver\cite{r10}. In contrast to traditional fixed position antennas (FPAs), MAs offer flexibility to enhance the channel conditions and improve both communication and sensing performance.

\subsection{Related Works} 

As previously discussed, DoFs play a pivotal role in enabling the dual objectives of high-performance communication and precise sensing within ISAC systems. Consequently, numerous researchers have explored the integration of ISAC systems with advanced technologies, including MIMO\cite{r39}, \cite{r40}, mMIMO\cite{r41}, \cite{r42}, antenna selection\cite{r43}, \cite{r44}, and RIS\cite{r45}, \cite{r46}. In \cite{r39}, Liu \emph{et} al. proposed a beamforming design for a DFRC system in the mmWave band, where the high DoFs provided by MIMO technology enabled it to support joint sensing and communication tasks. Using the effects of joint beam tilting and beam splitting, Gao \emph{et} al. proposed an ISAC scheme for mMIMO systems in \cite{r41}, in which mMIMO and mmWave/terahertz (THz) technology were employed to increase transmission rates and sensing accuracy. To achieve an optimal performance trade-off between multi-user communication and radar sensing, Liu \emph{et} al. studied the transmit/receive antenna selection for ISAC systems in \cite{r43}, and jointly optimized antenna selection and transmit beamforming based on different requirements and available resources. To address the challenges posed by complex transmission environments, Liao \emph{et} al. proposed an RIS-assisted ISAC system design for cluttered scenarios in \cite{r45}, and RIS can offer significant DoFs to reconstruct or enhance channel characteristics.

Presently, MA is employed to change channel characteristics, enhance communication system performance, reduce energy consumption, and adapt to the complexities of wireless environments. In \cite{r3}, Zhu \emph{et} al. developed a field-response channel model for MAs by utilizing the amplitude, phase, and angle of arrival/departure (AoA/AoD) information for each multichannel path under far-field conditions. In order to unlock more of the potential, Ma \emph{et} al. introduced an MA-assisted MIMO communication system in \cite{r2}, which demonstrated a substantial increase in channel capacity over the FPA system by flexibly moving of the transmit and receive MAs. For physical layer security, Tang \emph{et} al. considered an MA-assisted secure MIMO communication system in \cite{r48}, which uses the MA equipped at the BS to provide system security. In \cite{r10}, Zhu \emph{et} al. explored MA-enhanced multiple access channels (MAC), where multiple users equipped with a single MA transmit signal to the BS with an array of FPAs. To evaluate the performance gains attributed to MAs in the MAC, they formulated a power minimization problem aimed at simultaneously optimizing the positions of the MAs, the transmit power of the users, and the receive combining matrix at the BS. To better exploit the adaptability of MA systems in complex wireless environments, Wang \emph{et} al. investigated the interference channel of MA-assisted MISO system in \cite{r23}, which reduced the total transmit power by utilizing additional design DoFs provided by MAs to enhance the desired signal and suppress interference. In \cite{r24}, Ding \emph{et} al. conducted a study on a full-duplex communication system, focusing on reducing self-interference and improving signal reception by dynamically adjusting the positions of the transmit and receive antennas.

 
In parallel, researchers have shifted their focus towards exploring the impacts of MA on sensing and even ISAC. The multi-dimensional DoFs offered by MA hold significant potential for advancing sensing and communication functionalities. By dynamically adjusting the spatial positions of antennas, MA can enhance beamforming gain, mitigate complex interference environments, and achieve performance levels unattainable with FPA. In \cite{r25}, Ma \emph{et} al. conducted a study on multi-beamforming using linear MA arrays, capitalizing on the new DoFs afforded by moving  the antenna positions. They concluded that the multi-beamforming design using MA arrays significantly outperforms conventional FPA arrays in terms of beamforming gain and interference suppression, providing a reference for the application to sensing. Based on this, Ma \emph{et} al. introduced a wireless sensing system equipped with MAs that dynamically adjusted the positions of antenna elements in \cite{r26}. This system significantly enhanced the sensing performance over traditional FPA arrays. In \cite{r27} and \cite{r49}, Lyu \emph{et} al. explored a flexible beamforming design for ISAC systems utilizing MAs to dynamically adjust beam directions. Specifically, the BS transmitter (Tx) is equipped with one-dimensional MAs, and the receiver (Rx) with conventional FPAs. This configuration enhances the dual capabilities of sensing and communication through strategic antenna position adjustments. However, the adoption of one-dimensional MAs may result in untapped full potential for ISAC. To this end, Wu \emph{et} al. investigated the RIS-assisted ISAC system in \cite{r28}, where Tx deploys two-dimensional MAs, and assumed that between Tx and target only exists the reflective link of RIS. Then, they significantly improved the performance of the dual-function system by optimizing the Tx beamforming and the DoFs provided by the RIS and MAs.

\subsection{Motivation and Contributions}

To thoroughly explore the enhancement potential of MA for ISAC systems, we propose the implementation of a two-dimensional MA-assisted ISAC system, wherein MAs are deployed at both the Tx and Rx. The contributions of our work are summarized as follows:

\begin{itemize} 
	\item We develop a novel optimization framework for MA-assisted ISAC systems that leverages the additional spatial DoFs provided by 2D MAs. The framework jointly optimizes receive beamforming vectors, transmit beamforming matrices, and antenna positions to maximize the sensing signal-to-interference-plus-noise ratio (SINR) while guaranteeing communication SINR of multiple users simultaneously.
	\item To address the highly nonconvex nature of the optimization problem, we introduce a successive convex approximation (SCA) algorithm based on alternating optimization (AO) techniques. Specifically, the problem is decomposed into two primary types: beamforming and antenna position optimization. We derive a closed expression for the received beamforming vector for the radar echo and obtain the transmit beamforming matrix by the Charnes-Cooper transformation and semidefinite relaxation (SDR) method, and prove that the transmit matrix is the rank-one solution. Then, we obtain the solutions for the Tx and Rx antenna positions by the second-order Taylor expansion and SCA. Finally, we analyze the convergence and complexity of the proposed algorithm.
	\item Simulation results demonstrate that MAs can achieve substantial gains at both the Tx and Rx ends, significantly enhancing the sensing performance of the system while ensuring communication quality. Furthermore, the additional DoFs provided by MAs allow for a comprehensive performance improvement of the ISAC system, resulting in reduced resource needs, including fewer antennas, RF links, and lower power consumption.
	Simulation results validate the effectiveness of the proposed framework and algorithm. Compared to baseline schemes, the MA-assisted ISAC system achieves significant gains in sensing performance and communication quality, with over 50\% improvement in sensing SINR while requiring fewer antennas, RF links, and lower power consumption.

\end{itemize}

  \begin{figure}
 	\centering
 	\includegraphics[width=0.469\textwidth,angle=0]{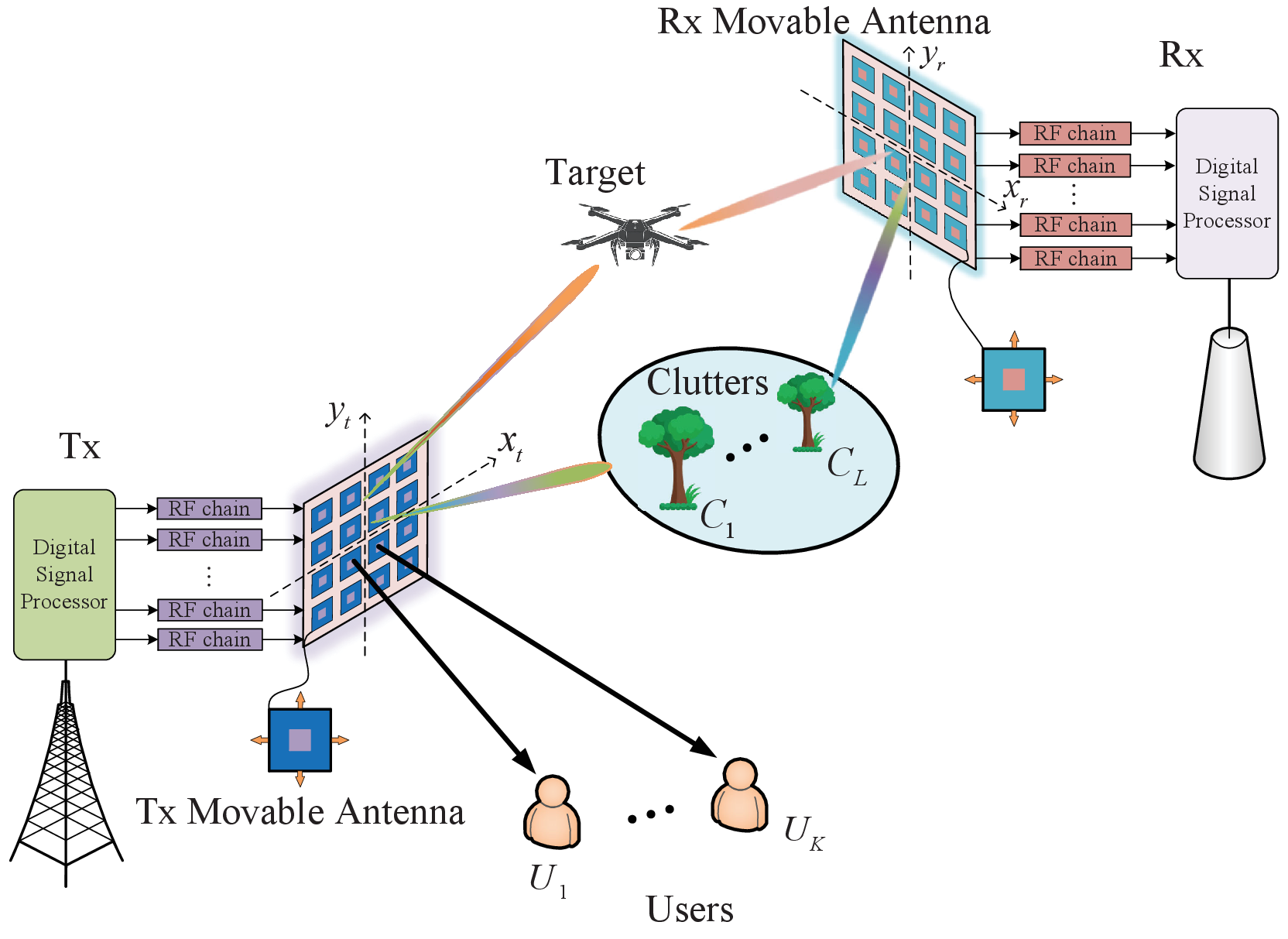}
 	\centering
 	\caption{The system model illustrates an MA-assisted ISAC framework. At the Tx and Rx, MAs dynamically adjust their 2D positions to optimize communication with users while simultaneously sensing a target amidst clutters. The flexible MAs provide additional spatial DoFs.}
 	\label{Fig_1}
 \end{figure}

\subsection{Organization}
The structure of this paper is as follows: section II analyzes the model of the MA-assisted ISAC system and formulates the corresponding optimization problem. Section III details the joint beamforming and MA positions optimization scheme. Section IV presents the simulation results. Finally, section VI offers the conclusions.

\emph{Notations}: ${\left(  \cdot  \right)^*}$ and ${\left(  \cdot  \right)^H}$ are the conjugate and conjugate transpose, respectively. $\left|  \cdot  \right|$ and $\left\|  \cdot  \right\|_F$ represent the absolute value of scalar and $F$-norm of complex vector, respectively. ${\mathbb{C}^{a \times b}}$ and ${\mathbb{R}^{a \times b}}$ denote $a \times b$ complex and real matrices, respectively. $x \sim {\cal C}{\cal N}\left( {u,v} \right)$ represents that the complex Gaussian variable with mean $u$ and variance $v$. ${\mathfrak {R}}\left( {{s}} \right)$ is the real part of the complex scalar $s$, respectively. The phase of complex value $s$ is denoted by $\angle s$. 





\section{System Model and problem formulation}


As shown in Fig. \ref{Fig_1}, we consider an ISAC system comprising a pair of transceiver BSs\footnote{To avoid self-interference, we use a bistatic system with separated Tx and Rx \cite{r1}.}, $K$ users, and a point-like target\footnote{Although this paper focuses on a single target, the proposed problem can be extended to multiple targets. With some modifications, the proposed algorithm can be applied to multiple target scenarios.}. Specifically, $N$ transmit and $M$ receive MAs, deployed at the Tx and Rx, respectively, are all configured as planar arrays (PAs). For simplicity, we denote ${\cal K} = \left\{ {1,2, \ldots, K} \right\}$, ${\cal N} = \left\{ {1,2, \ldots, N} \right\}$, and ${\cal M} = \left\{ {1,2, \ldots, M} \right\}$. Each user is considered for deployment of a conventional FPA. The MA modules are connected to the RF chain via flexible cables, allowing for real-time positional adjustments. The spatial positions of the $n$-th transmit MA and the $m$-th receive MA are denoted using the Cartesian coordinates ${{\bf{t}}_{{n}}} = {\left[ {{x_{{n}}},{y_{{n}}} } \right]^T} \in {{\cal C}_t}$ and ${{\bf{r}}_{{m}}} = {\left[ {{x_{{m}}},{y_{{m}}} } \right]^T} \in {{\cal C}_r}$, respectively. Herein, ${{\cal C}_t}$ and ${{\cal C}_r}$ represent specified two-dimensional (2D) regions where the transmit and receive MAs are capable of unrestricted movement, respectively. In this paper, we consider narrowband quasi-static channels, assuming that MA can move at a reasonable speed, rendering the overhead for adjusting MA positions negligible relative to the extended channel coherence time\cite{r2}. Furthermore, it is important to note that the channel characteristics associated with BS vary as the location of MAs is altered. We represent the positions of the transmit and receive MAs by the coordinates ${\bf{t}} = \left[ {{{\bf{t}}_1},{{\bf{t}}_2}, \ldots ,{{\bf{t}}_{{N}}}} \right] \in {{\mathbb R}^{2 \times {N}}}$ and ${\bf{r}} = \left[ {{{\bf{r}}_1},{{\bf{r}}_2}, \ldots ,{{\bf{r}}_{{M}}}} \right] \in {{\mathbb R}^{2 \times {M}}}$, respectively.

\subsection{Channel Model}
The channel vector between the user/target and the BS is influenced by both the propagation environment and the position of MA. Given that the antenna displacement is negligible in comparison to the signal transmission distance, we adopt the far-field model\cite{r10}. Based on this assumption, the plane wave model can be established from the PA of the BS for each user and target. Meanwhile, the AoD, AoA, and the amplitude of the complex coefficients between the BS and each node remain invariant, while only the phase of the channel path varies with the position of MA\cite{r3}. Therefore, we employ the field response-based channel model as described in \cite{r3}, wherein the channel response is characterized as a superposition of the signal coefficients from the multiple propagation paths between transceivers. 

Initially, we consider the communication channel and denote $L_k^t,k = \left\{ {1,2, \ldots ,K} \right\}$ as the number of propagation paths from the BS to user $k$. The elevation and azimuth AoDs of the $p$-th transmission path between BS and user $k$ are denoted as $\phi _{k,p}^t  \in \left[ {-\pi/2,\pi/2 } \right] $ and $\theta _{k,p}^t  \in \left[ {-\pi/2,\pi/2 } \right]$, $1 \le p \le L_k^t$, respectively. Then, the signal propagation phase difference between the $p$-th path of user $k$ and the reference point is denoted by 
\begin{equation}\label{eqn1}
	\rho _{k,p}^t\left( {{{\bf{t}}_{{n}}}} \right) = {x_{{n}}}\sin \phi _{k,p}^t\cos \theta _{k,p}^t + {y_{{n}}}\cos \phi _{k,p}^t.
\end{equation}

Consequently, the field response vector (FRV) of the transmission channel path between the $n$-th MA and user $k$ at the BS is denoted by 

\begin{equation}\label{eqn2}
	{{\bf{f}}_{k}}\left( {{{\bf{t}}_n}} \right) \!=\!\! {\left[ {{e^{j\frac{{2\pi }}{\lambda }\rho _{k,1}^t\left( {{{\bf{t}}_n}} \right)}},{e^{j\frac{{2\pi }}{\lambda }\rho _{k,2}^t\left( {{{\bf{t}}_n}} \right)}}, \ldots ,{e^{j\frac{{2\pi }}{\lambda }\rho _{k,L_k^t}^t\left( {{{\bf{t}}_n}} \right)}}} \right]^T}\!\!,
\end{equation}where $\lambda $ denotes the wavelength. Then, the field response matrix (FRM) between the BS and the user $k$ is denoted as 

\begin{equation}\label{eqn3}
	{{\bf{F}}_{k}}\left( {\bf{t}} \right) = \left[ {{{\bf{f}}_{k}}\left( {{{\bf{t}}}_1} \right),{{\bf{f}}_{k}}\left( {{{\bf{t}}_2}} \right), \ldots ,{{\bf{f}}_{k}}\left( {{{\bf{t}}_{{N}}}} \right)} \right] \in {{\mathbb C}^{L_k^t \times {N}}}.
\end{equation}Additionally, the path response vector (PRV), which represents the multipath response coefficient from the reference point in the reception area to user $k$, is denoted by 

\begin{equation}\label{eqn4}
	{{\bf{g}}_{k}} = {\left[ {{g_{k,1}},{g_{k,2}}, \ldots ,{g_{k,L_k^t}}} \right]^T} \in {{\mathbb C}^{L_k^t}}.
\end{equation}Then, the channel vector between the BS and user $k$ is obtained as 

\begin{equation}\label{eqn5}
	{{\bf{h}}_{k}}\left( {\bf{t}} \right) = {\bf{F}}_{k}^H\left( {\bf{t}} \right){{\bf{g}}_{k}}.
\end{equation}

Similarly, the channel matrices between the Tx and the target/clutters are denoted by

\begin{equation}\label{eqn6}
	{{\bf{h}}_{t,q}}\left( {\bf{t}} \right) = {\bf{F}}_{t,q}^H\left( {\bf{t}} \right){{\bf{g}}_{t,q}},
\end{equation}where $q \in \left\{ {d,l} \right\}$, $1 \le l \le L$, $d$ and $l$ denote the target and clutter, respectively. Likewise, for the receiving channel of the Rx, we first define the signal propagation phase difference between the $p$-th path of target $q$ and the reference point as 
 
\begin{equation}\label{eqn7}
   \rho _{q,p}^r\left( {{{\bf{r}}_{{m}}}} \right) = {x_{{m}}}\sin \phi _{q,p}^r\cos \theta _{q,p}^r + {y_{{m}}}\cos \phi _{q,p}^r,
\end{equation}where $\phi _{q,p}^r  \in \left[ {0,\pi } \right] $ and $\theta _{q,p}^r  \in \left[ {0,\pi } \right]$, $1 \le p \le L_q^r$ are elevation and azimuth AoAs, respectively, and $L_q^r$ is the number of propagation paths. Then, the FRVs of the transmission channel paths between target/clutters and the $m$-th MA at the Rx are denoted by

\begin{equation}\label{eqn8}
	{{\bf{f}}_{r,q}}\left( {{{\bf{r}}_m}} \right) \!=\!\! {\left[ {{e^{j\frac{{2\pi }}{\lambda }\rho _{q,1}^r\left(\! {{{\bf{r}}_m}} \!\right)}},{e^{j\frac{{2\pi }}{\lambda }\rho _{q,2}^r\left(\! {{{\bf{r}}_m}} \!\right)}}, \ldots ,{e^{j\frac{{2\pi }}{\lambda }\rho _{q,L_q^r}^r\left( \!{{{\bf{r}}_m}} \!\right)}}} \right]^T}\!\!.
\end{equation}Thus, the FRMs between the BS and target/clutters are represented as 

\begin{equation}\label{eqn9}
{{\bf{F}}_{r,q}}\left( {\bf{r}} \right) = \left[ {{{\bf{f}}_{r,q}}\left( {{{\bf{r}}_1}} \right),{{\bf{f}}_{r,q}}\left( {{{\bf{r}}_2}} \right), \ldots ,{{\bf{f}}_{r,q}}\left( {{{\bf{r}}_{{M}}}} \right)} \right] \in {{\mathbb C}^{L_q^r \times {M}}}.
\end{equation}The PRVs between the reference point and target/clutters is denoted by 

\begin{equation}\label{eqn10}
{{\bf{g}}_{r,q}} = {\left[ {{g_{q,1}},{g_{q,2}}, \ldots ,{g_{q,L_q^r}}} \right]^T} \in {{\mathbb C}^{L_q^r}}.
\end{equation}

Therefore, the received channel vector between the Rx and the target $q$ is obtained as 
 
\begin{equation}\label{eqn11}
	{{\bf{h}}_{r,q}}\left( {\bf{r}} \right) = {\bf{F}}_{r,q}^H\left( {\bf{r}} \right){{\bf{g}}_{r,q}}.
\end{equation}

 \subsection{Signal Model}
The transmit signal can be expressed as

\begin{equation}\label{eqn12}
	{\bf{s}} = \sum\nolimits_{k = 1}^K {{{\bf{w}}_k}{s_{c,k}}}  + \sum\nolimits_{n = K + 1}^N {{{\bf{w}}_n}{s_{r,n}}},
\end{equation}where ${{s_{c,k}}}$ is the communication signal received by user $k$ and ${{s_{r,n}}}$ is the dedicated radar signal, with ${\mathbb E}\left\{ {\left| {{s_{c,k}}} \right|}^2 \right\} = 1$ and ${\mathbb E}\left\{ {\left| {{s_{r,n}}} \right|}^2 \right\} = 1$ being satisfied. The communication and radar signals are assumed to exhibit statistical independence and do not have correlation\cite{r37}, i.e., ${\mathbb E}\left\{ {\left| {{s_{c,k}}s_{r,n}^H} \right|} \right\} = 0$, $\forall k,n$. Furthermore, ${{\bf{w}}_k} \in {{\mathbb C}^{N}}$ and ${{\bf{w}}_n} \in {{\mathbb C}^{N}}$ are the beamforming vectors for communication and radar, respectively. Therefore, the signal received by user $k$ is represented as

\begin{equation}\label{eqn13}
  	{y_k} = {\bf{h}}_{k}^H\left( {\bf{t}} \right) 	{\bf{s}} + {n_k},
\end{equation}where ${n_k}  \sim \mathcal{CN} \left( {0,\sigma _k^2} \right)$ is the additive white Gaussian noise (AWGN) at user $k$. Therefore, the SINR of the user $k$ is  

\begin{equation}\label{eqn14}
	\gamma _k^c = \frac{{{{\left| {{\bf{h}}_k^H\left( {\bf{t}} \right){{\bf{w}}_k}} \right|}^2}}}{{\sum\nolimits_{n \ne k}^N {{{\left| {{\bf{h}}_k^H\left( {\bf{t}} \right){{\bf{w}}_n}} \right|}^2}}  + \sigma _k^2}}.
\end{equation}

Similarly, the radar echo received at the Rx, which is reflected from the target, is denoted by 

\begin{equation}\label{eqn15}
	{{\bf{y}}_r} = \sum\nolimits_{q=1}^{L + 1} {{\alpha _q}{{\bf{h}}_{r,q}}\left( {\bf{r}} \right){\bf{h}}_{t,q}^H\left( {\bf{t}} \right){\bf{s}}}  + {{\bf{n}}_r},
\end{equation}where ${\alpha _q} \sim \mathcal{CN} \left( {0,\nu^2} \right)$ represents the complex-valued radar cross-section coefficient of the target $q$. ${{\bf{n}}_r}  \sim \mathcal{CN} \left( {0,\sigma _r^2{\bf{I}}} \right)$ is the AWGN at the Rx\cite{r37}. To enhance the sensing capability, the receive beamforming is utilized, and thus the output signal is  
\begin{equation}\label{eqn16}
	\begin{array}{l}
		y_r = {\bf{u}}_r^H{{\bf{y}}_r} \\
		\;\;\;\;\;      = \sum\nolimits_{q=1}^{L + 1} {{\alpha _q}{\bf{u}}_r^H{{\bf{h}}_{r,q}}\left( {\bf{r}} \right){\bf{h}}_{t,q}^H\left( {\bf{t}} \right){\bf{s}}}  + {\bf{u}}_r^H{{\bf{n}}_r}.
	\end{array}
\end{equation}

Then, the received sensing SINR of the target can be expressed as

\begin{equation}\label{eqn17}
    {\gamma _r} = \frac{{\sum\nolimits_{n = 1}^N {{{\left| {{\alpha _d}{\bf{u}}_r^H{{\bf{H}}_d}\left( {{\bf{r}},{\bf{t}}} \right){{\bf{w}}_n}} \right|}^2}} }}{{\sum\nolimits_{l = 1}^L {\sum\nolimits_{n = 1}^N {{{\left| {{\alpha _{{l}}}{\bf{u}}_r^H{{\bf{H}}_{{l}}}\left( {{\bf{r}},{\bf{t}}} \right){{\bf{w}}_n}} \right|}^2}} }  + \sigma _r^2{\left\| {{{\bf{u}}_r}} \right\|^2}    }},
\end{equation}where ${{\bf{H}}_q}\left( {{\bf{r}},{\bf{t}}} \right) = {{\bf{h}}_{r,q}}\left( {\bf{r}} \right){\bf{h}}_{t,q}^H\left( {\bf{t}} \right)$, $q \in \left\{ {d,{l}} \right\}$.

\subsection{Problem Formulation}
In this subsection, we explore the design of beamforming and the optimization of the MA positions at the BS. We account for the presence of $L$ clutter targets in this ISAC system, whose radar echoes interfere with the Rx. Furthermore, the locations of the target and clutters are determined during the previous detection phase, enabling their utilization in the tracking phase. Thus, we assume that the complete CSI of each target is fully known at the BS. Additionally, the user's CSI is also assumed to be perfectly obtainable using the corresponding channel estimation method\cite{r4}, \cite{r5}. Subsequently, the joint beamforming and MAs position optimization problem aims to maximize the sensing SINR of the desired target while satisfying the downlink communication requirements, which can be formulated as 

\begin{align}
	&	P1:{\rm{   }} \mathop {\max }\limits_{\left\{ {{{\bf{w}}_n}} \right\}_{n = 1}^N,{{\bf{u}}_r},{\bf{t}},{\bf{r}}} 	{\gamma _r} \label{eqn18}\\
	&\;\;\;\;\;\;\;\;\;    {\rm{  s.t. }} \;\gamma _k^c \ge \gamma _k^{th}, k \in  {\cal K}, \tag{\ref{eqn18}{a}}\\
	&\;\;\;\;\;\;\;\;\;\;\;\;\;\;\;  {{\bf{t}}_n} \in {{\cal C}_t}, n \in  {\cal N}, \tag{\ref{eqn18}{b}}\\
	&\;\;\;\;\;\;\;\;\;\;\;\;\;\;\;  {\left\| {{{\bf{t}}_n} - {{\bf{t}}_a}} \right\|_2} \ge {D}, a \in {{\cal N}_{ - n}}, \tag{\ref{eqn18}{c}}\\
	&\;\;\;\;\;\;\;\;\;\;\;\;\;\;\;  {{\bf{r}}_m} \in {{\cal C}_r}, m \in  {\cal M}, \tag{\ref{eqn18}{d}}\\
	&\;\;\;\;\;\;\;\;\;\;\;\;\;\;\;  {\left\| {{{\bf{r}}_m} - {{\bf{r}}_b}} \right\|_2} \ge {D}, b \in {{\cal M}_{ - m}}, \tag{\ref{eqn18}{e}}\\
	&\;\;\;\;\;\;\;\;\;\;\;\;\;\;\;  \sum\nolimits_{n = 1}^N {{{\left\| {{{\bf{w}}_n}} \right\|}^2}}  \le {P_{th}}, \tag{\ref{eqn18}{f}}
\end{align}where $\gamma _k^{th}$ is the communication SINR threshold for user $k$, $D$ is the minimum distance between MAs, and ${P_{th}}$ is the maximum transmit power. The sets ${{\cal N}_{ - n}}$ and ${{\cal M}_{ - m}}$ represent the elements excluding $n$ in set ${\cal N}$ and $m$ in set ${\cal M}$, respectively. Evidently, $P$1 poses significant challenges due to the nonconcave nature of the objective function and constraints (\ref{eqn18}a) with respect to ${\bf{t}}$ and ${\bf{r}}$, as well as the minimum distance constraints (\ref{eqn18}c) and (\ref{eqn18}e). Moreover, the coupling among $\left\{ {{{\bf{w}}_n}} \right\}_{n = 1}^N$, ${{\bf{u}}_r}$, ${\bf{t}}$, and ${\bf{r}}$ further complicates the resolution of $P$1. Consequently, standard convex optimization tools are inadequate for solving this problem.

\section{Joint beamforming and the MA positions optimization}

In this section, we develop an AO algorithm that iteratively determines the optimal solution for each individual variable. Specifically, we optimize multiple subproblems in turn by holding the remaining variables constant.

\subsection{Receive Beamforming Optimization}


Given $\left\{ {{{\bf{w}}_n}} \right\}_{n = 1}^N$, ${\bf{t}}$, and ${\bf{r}}$, we initially optimize the received beamformer ${{\bf{u}}_r}$. Clearly, ${{\bf{u}}_r}$ is included solely in the objective function, so $P$1 simplifies to 

\begin{equation}\label{eqn19}
 P2:\mathop {\max }\limits_{{{\bf{u}}_r}} \frac{{{{\left| {{\bf{u}}_r^H{\bf{d}}{{_d^{\prime}} }} \right|}^2}}}{{{\bf{u}}_r^H\left( {     \sum\nolimits_{l = 1}^L {{\bf{R}}{{_{l}^{\prime}}}  } + \sigma _r^2{\bf{I}}} \right){\bf{u}}_r^{}}},
\end{equation}where ${\bf{R}}{_{l}^{\prime} } = {{\left| {{\alpha _{{l}}}} \right|}^2} \sum\nolimits_{n = 1}^N {{{\bf{H}}_{{l}}}\left( {{\bf{r}},{\bf{t}}} \right){{\bf{w}}_n}{\bf{w}}_n^H{\bf{H}}_{{l}}^H\left( {{\bf{r}},{\bf{t}}} \right)}$, and ${\bf{d}}{_d^{\prime} } = {{\left| {{\alpha _t}} \right|}^2} \sum\nolimits_{n = 1}^N {{{\bf{H}}_d}\left( {{\bf{r}},{\bf{t}}} \right){{\bf{w}}_n}}$. Given that $P$2 is a classical SINR maximization problem, which can be equivalently solved by addressing a minimum variance distortion-free response (MVDR) problem\cite{r6},\cite{r7}, we provide its closed-form solution as

\begin{equation}\label{eqn20}
	{\bf{u}}_r^ *  = \frac{{{{\left( {{\sum\nolimits_{l = 1}^L {{\bf{R}}{{_{l}^{\prime}}}  }} + \sigma _r^2{\bf{I}}} \right)}^{ - 1}} {\bf{d}}{_d^{\prime} }}}{{  {{\bf{d}}{_d^{\prime} }}^H {{\left( {\sum\nolimits_{l = 1}^L {{\bf{R}}{{_{l}^{\prime}}}  } + \sigma _r^2{\bf{I}}} \right)}^{ - 1}} {\bf{d}}{_d^{\prime} }}}.
\end{equation}

\subsection{Transmit Beamforming Optimization}
 

After updating ${{\bf{u}}_r}$ and fixing ${\bf{t}}$ and ${\bf{r}}$, the transmit beamforming matrix will be optimized. To eliminate the quadratic forms, we first introduce the following definitions: ${{\bf{W}}_n} = {{\bf{w}}_n}{\bf{w}}_n^H$, ${\widetilde {\bf{H}}_q} = {\bf{H}}_q^H\left( {{\bf{r}},{\bf{t}}} \right){\bf{u}}_r^{}{\bf{u}}_r^H{{\bf{H}}_q}\left( {{\bf{r}},{\bf{t}}} \right)$, and ${\widehat {\bf{H}}_n} = {{\bf{h}}_n}\left( {\bf{t}} \right){\bf{h}}_n^H\left( {\bf{t}} \right)$, where $q \in \left\{ {d,{l}} \right\}$. Then, $P$1 is reformulated as

\begin{align}
	&	P3:{\rm{   }} \mathop {\max }\limits_{\left\{ {{{\bf{w}}_n}} \right\}_{n = 1}^N } 	\frac{{{{\left| {{\alpha _t}} \right|}^2}\sum\nolimits_{n = 1}^N {{\rm{tr}}\left( {{{\widetilde {\bf{H}}}_d}{{\bf{W}}_n}} \right)} }}{{\sum\nolimits_{l = 1}^L {{{\left| {{\alpha _{{l}}}} \right|}^2}\sum\nolimits_{n = 1}^N {{\rm{tr}}\left( {{{\widetilde {\bf{H}}}_{{l}}}{{\bf{W}}_n}} \right)} }  + \sigma _r^2 }} \label{eqn21}\\
	&\;\;\;\;\;\;\;\;\;    {\rm{  s.t. }} \;\frac{{{\rm{tr}}\left( {{{\widehat {\bf{H}}}_k}{{\bf{W}}_k}} \right)}}{{\sum\nolimits_{n \ne k}^N {{\rm{tr}}\left( {{{\widehat {\bf{H}}}_k}{{\bf{W}}_n}} \right)}  + \delta _k^2}} \ge \gamma _k^{th},  k \in  {\cal K}, \tag{\ref{eqn21}{a}}\\
	&\;\;\;\;\;\;\;\;\;\;\;\;\;\;\;  \sum\nolimits_{n = 1}^N {{\rm{tr}}\left( {{{\bf{W}}_n}} \right)}  \le {P_{th}}, \tag{\ref{eqn21}{b}}\\
	&\;\;\;\;\;\;\;\;\;\;\;\;\;\;\;  {\rm{rank}}\left( {{{\bf{W}}_n}} \right) = 1, n \in  {\cal N}, \tag{\ref{eqn21}{c}}\\
	&\;\;\;\;\;\;\;\;\;\;\;\;\;\;\;  {{\bf{W}}_n} \succeq 0,  n \in  {\cal N}. \tag{\ref{eqn21}{d}}
\end{align}

To address the nonconvexity of the objective function, we utilize the Charnes-Cooper transformation by first letting $\ell  = {\left( {\sum\nolimits_{l = 1}^L {{{\left| {{\alpha _{{l}}}} \right|}^2}\sum\nolimits_{n = 1}^N {{\rm{tr}}\left( {{{\widetilde {\bf{H}}}_{{l}}}{{\bf{W}}_n}} \right)} }  + \sigma _r^2} \right)^{ - 1} }$ and ${{\bf{X}}_n} = \ell {{\bf{W}}_n}$. Moreover, we adopt the SDR to relax the rank-1 constraint (\ref{eqn21}c). After the above transformation, $P$3 is rewritten as

\begin{align}
	&	P3.1:\mathop {\max }\limits_{\left\{ {{{\bf{X}}_n}} \right\}_{n = 1}^N, \ell} {\left| {{\alpha _t}} \right|^2}\sum\nolimits_{n = 1}^N {{\rm{tr}}\left( {{{\widetilde {\bf{H}}}_d}{{\bf{X}}_n}} \right)}  \label{eqn22}\\
	&\;\;\;       {\rm{  s.t. }} \;   \sum\nolimits_{l = 1}^L {{{\left| {{\alpha _{{l}}}} \right|}^2}\sum\nolimits_{n = 1}^N {{\rm{tr}}\left( {{{\widetilde {\bf{H}}}_{{l}}}{{\bf{X}}_n}} \right)} }  + \sigma _r^2\ell  = 1, \tag{\ref{eqn22}{a}}\\
	&\;\;\;\;\;\;\;\;\;   {\rm{tr}}\left( {{{\widehat {\bf{H}}}_k}{{\bf{X}}_k}} \right) - \gamma _k^{th}\sum\nolimits_{n \ne k}^N {{\rm{tr}}\left( {{{\widehat {\bf{H}}}_k}{{\bf{X}}_n}} \right)} \nonumber \\
	&\;\;\;\;\;\;\;\;\;\;\;\;\;\;\;\;\;\;\;\;\;\;\;\;\;\;\;    - \ell \gamma _k^{th}\delta _k^2 \ge 0, k \in  {\cal K}, \tag{\ref{eqn22}{b}}\\
	&\;\;\;\;\;\;\;\;\;   \sum\nolimits_{n = 1}^N {{\rm{tr}}\left( {{{\bf{X}}_n}} \right)}  \le {\rm{ }}\ell {P_{th}}, \tag{\ref{eqn22}{c}} \\
	&\;\;\;\;\;\;\;\;\;   {{\bf{X}}_n} \succeq 0,\ell  \ge 0, n \in  {\cal N}. \tag{\ref{eqn22}{d}} 
\end{align}

Up to this point, $P$3.1 is convex and solvable, and its can be solved according to the existing CVX tool. Then, we can obtain ${\bf{W}}_n^{\star} = {\bf{X}}_n^{}/\ell $. Since SDR is employed, the following proposition and theorem are given to prove that the rank-1 relaxation is tight, which guarantees that ${\bf{W}}_n^{\star}$ can be obtained by singular value decomposition (SVD).

\emph{Proposition 1: Assuming that $\gamma^{\star} $ is the optimal value of $P$3.1, an equivalent power minimization problem is given as follows:}

\begin{align}
	& P3.2:\mathop {\min }\limits_{\left\{ {{{\bf{X}}_n}} \right\}_{n = 1}^N,\ell } \;\sum\nolimits_{n = 1}^N {{\rm{tr}}\left( {{{\bf{X}}_n}} \right)}     \label{eqn23}\\
	&\;\;\;\;\;\;\;\;\;   {\rm{  s.t. }} \; {\left| {{\alpha _t}} \right|^2}\sum\nolimits_{n = 1}^N {{\rm{tr}}\left( {{{\widetilde {\bf{H}}}_d}{{\bf{X}}_n}} \right)}  \ge {\gamma ^ \star },  \tag{\ref{eqn23}{a}}\\
	&\;\;\;\;\;\;\;\;\;\;\;\;\;\;\;  (\ref{eqn22}a), (\ref{eqn22}b), (\ref{eqn22}d)  \tag{\ref{eqn23}{b}},
\end{align}\emph{where any feasible solution to $P$3.2 is an optimal solution to $P$3.1.}

\emph{Proof:} We assume that $\left\{ {{\bf{X}}_n^\circ  } \right\}_{n = 1}^N$ is optimal for $P$3.2, so ${\left| {{\alpha _t}} \right|^2}\sum\nolimits_{n = 1}^N {{\rm{tr}}\left( {{{\widetilde {\bf{H}}}_d}{\bf{X}}_n^ \circ } \right)}  \ge {\gamma ^ \star }$ holds due to (\ref{eqn23}a). Obviously, any feasible solution in $P$3.2 is optimal for $P$3.1, and its maximum value is reached at $\left\{ {{\bf{X}}_n^\circ  } \right\}_{n = 1}^N$. Thus, the optimal values for $P$3.2 are equivalently optimal for $P$3.1.$\hfill\blacksquare$

Based on Proposition 1, we can prove that the rank-1 condition in $P$3.1 by proving that in $P$3.2, and \emph{Theorem 1} is given.

\emph{Theorem 1: Any feasible solution to $P$3.2 satisfies $rank\left( {{\bf{X}}_n^ \circ } \right) = 1$, $1 \le n \le N$.}

\emph{Proof:} Please see the Appendix A.$\hfill\blacksquare$

\subsection{Receive Antenna Position Optimization}


Subsequent to the updates applied to ${{\bf{u}}_r}$ and $\left\{ {{{\bf{w}}_n}} \right\}_{n = 1}^N$, we optimize the parameter ${{{\bf{r}}}_m}$ with ${\bf{t}}$ and ${{{\bf{r}}}_b}$ $\left( { 1 \le b \ne m \le N } \right)$ held constant. Hence, we reformulate $P$1 while introducing the slack variables $\left\{ {\alpha ,\beta ,\chi } \right\}$, thereby simplifying it to $P$4.

\begin{align}
	&	P4:{\rm{   }} \mathop {\max }\limits_{ {\bf{r}}_m, \alpha ,\beta, \chi }\;\;	\chi  \label{eqn24}\\
	&\;\;\;\;\;\;\;\;\;    {\rm{  s.t. }} \; \alpha  \ge \beta \chi , \tag{\ref{eqn24}{a}}\\
	&\;\;\;\;\;\;\;\;\;\;\;\;\;\;\;  f\left( {{{\bf{r}}_m}} \right) \ge \alpha , \tag{\ref{eqn24}{b}}\\
	&\;\;\;\;\;\;\;\;\;\;\;\;\;\;\;  g\left( {{{\bf{r}}_m}} \right) \le \beta , \tag{\ref{eqn24}{c}}\\
	&\;\;\;\;\;\;\;\;\;\;\;\;\;\;\;  {{\bf{r}}_m} \in {{\cal C}_r} , \tag{\ref{eqn24}{d}}\\
	&\;\;\;\;\;\;\;\;\;\;\;\;\;\;\;  {\left\| {{{\bf{r}}_m} - {{\bf{r}}_b}} \right\|_2} \ge {D},b \in {{\cal M}_{ - m}}, \tag{\ref{eqn24}{e}} 
\end{align}where $f\left( {{{\bf{r}}_m}} \right)$ and $g\left( {{{\bf{r}}_m}} \right)$ are specified in (\ref{eqn25}) and (\ref{eqn26}), respectively, located at the top of the next page. In them, ${x_{m,n}} = {u^*_m}{\bf{f}}_{r,d}^H\left( {{{\bf{r}}_m}} \right){\widetilde {\bf{g}}_{d,n}}$, ${y_{m,n,l}} = {u^*_m}{\bf{f}}_{{r,l}}^H\left( {{{\bf{r}}_m}} \right){\widetilde {\bf{g}}_{{l},n}}$, and ${\widetilde {\bf{g}}_{q,n}} = {{\bf{g}}_{r,q}}{\bf{h}}_{t,q}^H\left( {\bf{t}} \right){{\bf{w}}_n}$, $q \in \left\{ {d,{l}} \right\}$. In both (\ref{eqn25}) and (\ref{eqn26}), the third term proves to be irrelevant for ${{{\bf{r}}}_m}$. Additionally, we provide a detailed derivation of the relevant terms ${\widetilde f\left( {{{\bf{r}}_m}} \right)}$ and ${\widetilde g\left( {{{\bf{r}}_m}} \right)}$ with respect to the variables in (\ref{eqn27}) and (\ref{eqn28}) at the top of the next page. For simplicity, we have

\begin{figure*}[ht]
	\begin{small}
	\begin{equation}\label{eqn25}
		f\left( {{{\bf{r}}_m}} \right) = \underbrace {{{\left| {{\alpha _d}} \right|}^2}\sum\limits_{n = 1}^N {\left( {{{\left| {{x_{m,n}}} \right|}^2} + 2\Re \left( {\sum\limits_{j = 1,j \ne m}^M {{x_{m,n}}x_{j,n}^*} } \right)}\!\!\! \right)} }_{\widetilde f\left( {{{\bf{r}}_m}} \right)} + {\left| {{\alpha _d}} \right|^2}\sum\limits_{n = 1}^N {\sum\limits_{j = 1,j \ne m}^M {\left( {{{\left| {{x_{j,n}}} \right|}^2} + 2\Re \left( {\sum\limits_{{\widetilde j} = j + 1,{\widetilde j} \ne m}^M { {{x_{{\widetilde j},n}}x_{j,n}^*} } } \right)}\!\!\! \right)} } .
	\end{equation}
    \end{small}
	\begin{small}
	\begin{equation}\label{eqn26}
		g\left( {{{\bf{r}}_m}} \right) = \underbrace {\sum\limits_{l = 1}^L \!{{{\left| {{\alpha _l}} \right|}^2}\!\sum\limits_{n = 1}^N\! {\left( \!\!{{{\left| {{y_{m,n,l}}} \right|}^2} \!+\! 2\Re \left(\! {\sum\limits_{j = 1,j \ne m}^M \!\!\! {{y_{m,n,l}}y_{j,n,l}^*} } \right)} \!\!\! \right)} } }_{\widetilde g\left( {{{\bf{r}}_m}} \right)} \!+\! \sum\limits_{l = 1}^L {{{\left| {{\alpha _l}} \right|}^2}\sum\limits_{n = 1}^N {\sum\limits_{j = 1,j \ne m}^M {\left( {{{\left| {{y_{j,n,l}}} \right|}^2} \!+\! 2\Re \left( {\sum\limits_{\widetilde j = j + 1,\widetilde j \ne m}^M {\!\!\!\!\! {{y_{\widetilde j,n,l}}y_{j,n,l}^*}  } }  \! \right)} \!\!\! \right)} } }  \!+\! \sigma _r^2.
	\end{equation}
    \end{small}
		\begin{equation}\label{eqn27} 
		 \widetilde f\left( {{{\bf{r}}_m}} \right) = {\left| {{\alpha _d}} \right|^2}\sum\limits_{n = 1}^N {\sum\limits_{j = 1}^{L_d^r} {\sum\limits_{p = 1}^{L_d^r} {\left[ {{{\left| {{u_m}} \right|}^2}\left| {{{\left[ {{{\widetilde {\bf{G}}}_{d,n}}} \right]}_{j,p}}} \right|\cos \lambda _{d,n,j,p}^m + 2\left| {{u_m}} \right|\left| {{u_b}} \right|\left| {{{\left[ {{{\widetilde {\bf{G}}}_{d,n}}} \right]}_{j,p}}} \right|\cos \lambda _{d,n,j,p}^b} \right]} } }, 
		\end{equation}
		\begin{equation}\label{eqn28}
			\widetilde g\left( {{{\bf{r}}_m}} \right) = \sum\limits_{l = 1}^L {{{\left| {{\alpha _l}} \right|}^2}\sum\limits_{n = 1}^N {\sum\limits_{j = 1}^{L_l^r} {\sum\limits_{p = 1}^{L_l^r} {\left[ {{{\left| {{u_m}} \right|}^2}\left| {{{\left[ {{{\widetilde {\bf{G}}}_{l,n}}} \right]}_{j,p}}} \right|\cos \lambda _{l,n,j,p}^m + \sum\limits_{b = 1,b \ne m}^M {2\left| {{u_m}} \right|\left| {{u_b}} \right|\left| {{{\left[ {{{\widetilde {\bf{G}}}_{l,n}}} \right]}_{j,p}}} \right|\cos \lambda _{l,n,j,p}^b} } \right]} } } } .
		\end{equation}
	\hrulefill
\end{figure*}

\begin{small}
	\begin{equation}\label{eqn100}
		\lambda _{q,n,j,p}^m = \frac{{2\pi }}{\lambda }\rho _{q,p}^r\left( {{{\bf{r}}_m}} \right) - \frac{{2\pi }}{\lambda }\rho _{q,j}^r\left( {{{\bf{r}}_m}} \right) + \angle {\left[ {{{\widetilde {\bf{G}}}_{q,n}}} \right]_{j,p}},
	\end{equation}
\end{small}
and
\begin{small}
	\begin{equation}\label{eqn101}
		\lambda _{q,n,j,p}^b = \frac{{2\pi }}{\lambda }\rho _{q,p}^r\left( {{{\bf{r}}_b}} \right) \!-\! \frac{{2\pi }}{\lambda }\rho _{q,j}^r\left( {{{\bf{r}}_m}} \right) \!+\! \angle {\left[ {{{\widetilde {\bf{G}}}_{q,n}}} \right]_{j,p}} \!\!\!\!\!- \angle {u_m} + \angle {u_b},
	\end{equation}
\end{small}where ${\widetilde {\bf{G}}_{q,n}} = {\widetilde {\bf{g}}_{q,n}}\widetilde {\bf{g}}_{q,n}^H$ and $q \in \left\{ {d,l} \right\}$.

Clearly, the problem remains nonconvex and intractable. To address this challenge, we employ the SCA method to transform the nonconvex constraints into a more tractable form. To take the lead, we deal with (\ref{eqn24}a), and using the arithmetic geometric mean inequality, we can derive its convex form as 

\begin{equation}\label{eqn29}
	\alpha  \ge \frac{1}{2}\left( {{\varsigma ^{\left( i \right)}}{\beta ^2} + {\chi ^2}/{\varsigma ^{\left( i \right)}}} \right),
\end{equation}where ${\varsigma ^{\left( i \right)}} = {\beta ^{\left( {i} \right)}}/{\chi ^{\left( {i} \right)}}$, which can be obtained in the $i$-th iteration.

Then, the second-order Taylor expansion is employed to reconstruct (\ref{eqn24}b) and (\ref{eqn24}c)\cite{r8}. By adjusting the Hessian matrix with positive or negative definite terms, we can ensure the concavity or convexity of functions, enabling the use of convex optimization techniques in non-convex problems. In addition, we  introduce two positive real numbers, ${\delta _f^i}$ and ${\delta _g^i}$, that satisfy $\delta _f^i{{\bf{I}}_2} \succeq {\nabla ^2}f\left( {{{\bf{r}}_m}} \right)$ and $\delta _g^i{{\bf{I}}_2} \succeq {\nabla ^2}g\left( {{{\bf{r}}_m}} \right)$, we establish a lower bound for $f\left( {{{\bf{r}}_m}} \right)$ and an upper bound for $g\left( {{{\bf{r}}_m}} \right)$ as follows:

\begin{equation}\label{eqn30}
	\begin{array}{l}
		f\left( {{{\bf{r}}_m}} \right) \ge f\left( {{\bf{r}}_m^{\left( i \right)}} \right) + \nabla f{\left( {{\bf{r}}_m^{\left( i \right)}} \right)^T}\left( {{{\bf{r}}_m} - {\bf{r}}_m^{\left( i \right)}} \right)\\
		\;\;\;\;\;\;\;\;\;\;\;\;\;\; - \frac{{\delta _f^i}}{2}{\left( {{{\bf{r}}_m} - {\bf{r}}_m^{\left( i \right)}} \right)^T}\left( {{{\bf{r}}_m} - {\bf{r}}_m^{\left( i \right)}} \right) \ge \alpha    ,
	\end{array}
\end{equation}
and
\begin{equation}\label{eqn31}
	\begin{array}{l}
		g\left( {{{\bf{r}}_m}} \right) \le g\left( {{\bf{r}}_m^{\left( i \right)}} \right) + \nabla g{\left( {{\bf{r}}_m^{\left( i \right)}} \right)^T}\left( {{{\bf{r}}_m} - {\bf{r}}_m^{\left( i \right)}} \right)\\
		\;\;\;\;\;\;\;\;\;\;\;\;\;\; + \frac{{\delta _g^i}}{2}{\left( {{{\bf{r}}_m} - {\bf{r}}_m^{\left( i \right)}} \right)^T}\left( {{{\bf{r}}_m} - {\bf{r}}_m^{\left( i \right)}} \right) \le \beta ,  
	\end{array}
\end{equation}where ${\bf{r}}_m^{\left( i \right)}$ is the local point obtained in the $i$-th iteration. The gradients 
 $\nabla f{\left( {{\bf{r}}_m } \right)}$ and $\nabla g{\left( {{\bf{r}}_m } \right)}$, along with the Hessian matrix ${\nabla ^2}f\left( {{{\bf{r}}_m}} \right)$ and ${\nabla ^2}g\left( {{{\bf{r}}_m}} \right)$, are detailed in Appendix B. Thus, we have
 
 \begin{small}
  \begin{equation}\label{eqn32}
 	\begin{array}{l}
 		\left\| {{\nabla ^2}f\left( {{{\bf{r}}_m}} \right)} \right\|_2^2 \!\le\!\! \left\| {{\nabla ^2}f\left( {{{\bf{r}}_m}} \right)} \right\|_F^2\!\! =\!\! {\left( {\frac{{{\partial ^2}f\left( {{{\bf{r}}_m}} \right)}}{{\partial x_m^2}}} \right)^2} \!\!\!+\!\! {\left( {\frac{{{\partial ^2}f\left( {{{\bf{r}}_m}} \right)}}{{\partial y_m^2}}} \right)^2}\\
 		\;\;\;\;\;\;\;\;\;\;\;\;\;\;\;\;\;\;\;\;\;\;\;\;\;\;\;\;\;\;\;\;\;\;\;\;\;\;\;\;\;\;\;    + {\left( {\frac{{{\partial ^2}f\left( {{{\bf{r}}_m}} \right)}}{{\partial {x_m}\partial {y_m}}}} \right)^2} \!\!\!+\!\! {\left( {\frac{{{\partial ^2}f\left( {{{\bf{r}}_m}} \right)}}{{\partial {y_m}\partial {x_m}}}} \right)^2}\\
 		\;\;\;\;\;\;\;\; \le 4\left( {\frac{{16{\pi ^2}}}{{{\lambda ^2}}}{{\left| {{\alpha _t}} \right|}^2}\sum\limits_{n = 1}^N {\sum\limits_{j = 1}^{L_d^r} {\sum\limits_{p = 1}^{L_d^r} {{{\left| {{u_m}} \right|}^2}\left| {{{\left[ {{{\widetilde {\bf{G}}}_{d,n}}} \right]}_{j,p}}} \right|} } } } \right.\\
 		\;\;\;\;\;\;\;\;  {\left. {\! +\! \frac{{8{\pi ^2}}}{{{\lambda ^2}}}{{\left| {{\alpha _t}} \right|}^2}\!\!\sum\limits_{n = 1}^N \!{\sum\limits_{b = 1,b \ne m}^M {\sum\limits_{j = 1}^{L_d^r} {\sum\limits_{p = 1}^{L_d^r} {\left| {{u_m}} \right|\!\!\left| {{u_b}} \right|\!\!\left|\! {{{\left[ {{{\widetilde {\bf{G}}}_{d,n}}} \right]}_{j,p}}} \right|} } } } } \!  \right)^2}
 	\end{array}
 \end{equation}
   \end{small}
 and 
 \begin{small}
 \begin{equation}\label{eqn33}
 	\begin{array}{l}
 		{\left\| {{\nabla ^2}g\left( {{{\bf{r}}_m}} \right)} \right\|^2_2} \!\le\!\! \left\| {{\nabla ^2}g\left( {{{\bf{r}}_m}} \right)} \right\|_F^2 \!\!=\!\! {\left( {\frac{{{\partial ^2}g\left( {{{\bf{r}}_m}} \right)}}{{\partial x_m^2}}} \right)^2} \!\!\!+\!\! {\left( {\frac{{{\partial ^2}g\left( {{{\bf{r}}_m}} \right)}}{{\partial y_m^2}}} \right)^2} \\
 		\;\;\;\;\;\;\;\;\;\;\;\;\;\;\;\;\;\;\;\;\;\;\;\;\;\;\;\;\;\;\;\;\;\;\;\;\;\;\;\;\;\;    + {\left( {\frac{{{\partial ^2}g\left( {{{\bf{r}}_m}} \right)}}{{\partial {x_m}\partial {y_m}}}} \right)^2} \!\!\!+\!\! {\left( {\frac{{{\partial ^2}g\left( {{{\bf{r}}_m}} \right)}}{{\partial {y_m}\partial {x_m}}}} \right)^2}\\
 		\;\;\;\;\;\;  \le 4\left( {\frac{{16{\pi ^2}}}{{{\lambda ^2}}}\sum\limits_{l = 1}^L {{{\left| {{\alpha _l}} \right|}^2}\sum\limits_{n = 1}^N {\sum\limits_{j = 1}^{L_l^r} {\sum\limits_{p = 1}^{L_l^r} {{{\left| {{u_m}} \right|}^2}\left| {{{\left[ {{{\widetilde {\bf{G}}}_{l,n}}} \right]}_{j,p}}} \right|} } } } } \right.\\
 	    \;\;\;\;\;\;  {\left. {\! +\! \frac{{4{\pi ^2}}}{{{\lambda ^2}}}\!\!\sum\limits_{l = 1}^L \!{{{\left| {{\alpha _l}} \right|}^2}\!\!\sum\limits_{n = 1}^N\! {\sum\limits_{b = 1,b \ne m}^M \!{\sum\limits_{j = 1}^{L_l^r} {\sum\limits_{p = 1}^{L_l^r} \!\!{\left| {{u_m}} \right|\!\!\left| {{u_b}} \right|\!\!\left|\!  {{{\left[ {{{\widetilde {\bf{G}}}_{l,n}}} \right]}_{j,p}}} \right|} } } } } } \!\right)^2}.
 	\end{array}
 \end{equation}
  \end{small}Then, due to $\left\| {{\nabla ^2}f\left( {{{\bf{r}}_m}} \right)} \right\|_2^2{{\bf{I}}_2} \succeq {\nabla ^2}f\left( {{{\bf{r}}_m}} \right)$ and $\left\| {{\nabla ^2}g\left( {{{\bf{r}}_m}} \right)} \right\|_2^2{{\bf{I}}_2} \succeq {\nabla ^2}g\left( {{{\bf{r}}_m}} \right)$, we select $\delta _f^i$ and $\delta _g^i$ as

  \begin{equation}\label{eqn34}
 	\begin{array}{l}
 		\delta _f^i = \frac{{32{\pi ^2}}}{{{\lambda ^2}}}{\left| {{\alpha _t}} \right|^2}\sum\limits_{n = 1}^N {\sum\limits_{j = 1}^{L_d^r} {\sum\limits_{p = 1}^{L_d^r} {{{\left| {{u_m}} \right|}^2}\left| {{{\left[ {{{\widetilde {\bf{G}}}_{d,n}}} \right]}_{j,p}}} \right|} } } \\
 		   + \frac{{8{\pi ^2}}}{{{\lambda ^2}}}{\left| {{\alpha _t}} \right|^2}\sum\limits_{n = 1}^N {\sum\limits_{b = 1,b \ne m}^M {\sum\limits_{j = 1}^{L_d^r} {\sum\limits_{p = 1}^{L_d^r} {\left| {{u_m}} \right|\left|\! {{u_b}} \right|\left|\! {{{\left[ {{{\widetilde {\bf{G}}}_{d,n}}} \right]}_{j,p}}}\! \right|} } } }, 
 	\end{array} 
 \end{equation}
and
  \begin{equation}\label{eqn35}
 	\begin{array}{l}
 		\delta _g^i = \frac{{32{\pi ^2}}}{{{\lambda ^2}}}\sum\limits_{l = 1}^L {{{\left| {{\alpha _l}} \right|}^2}\sum\limits_{n = 1}^N {\sum\limits_{j = 1}^{L_l^r} {\sum\limits_{p = 1}^{L_l^r} {{{\left| {{u_m}} \right|}^2}\left| {{{\left[ {{{\widetilde {\bf{G}}}_{l,n}}} \right]}_{j,p}}} \right|} } } } \\
 		 \!+\!\frac{{8{\pi ^2}}}{{{\lambda ^2}}}\! \sum\limits_{l = 1}^L {{{\left| {{\alpha _l}} \right|}^2}\sum\limits_{n = 1}^N {\sum\limits_{b = 1,b \ne m}^M {\sum\limits_{j = 1}^{L_l^r} {\sum\limits_{p = 1}^{L_l^r} {\left| {{u_m}} \right|\!\!\left| {{u_b}} \right|\!\left|\! {{{\left[ {{{\widetilde {\bf{G}}}_{l,n}}} \right]}_{j,p}}} \!\right|} } } } } .
 	\end{array}
 \end{equation}In this manner, by tightening $f\left( {{{\bf{r}}_m}} \right)$ in (\ref{eqn30}) to a lower bound and $g\left( {{{\bf{r}}_m}} \right)$ in (\ref{eqn31}) to an upper bound, we can iteratively approach a suboptimum point. Subsequently, we turn our goal to the last nonconvex constraint, as indicated in (\ref{eqn24}e). Given that ${\left\| {{{\bf{r}}_m} - {{\bf{r}}_j}} \right\|_2}$ is convex with respect to ${{\bf{r}}_m}$, we employ the first-order Taylor expansion at ${{\bf{r}}_m^{\left( i \right)}}$ to obtain  
 
\begin{small}
\begin{equation}\label{eqn36}
	\begin{array}{l}
		{\left\| {{{\bf{r}}_m} - {{\bf{r}}_b}} \right\|_2}\\
		\;\;\ge {\left\| {{\bf{r}}_m^{\left( i \right)} - {{\bf{r}}_b}} \right\|_2} + \nabla {\left( {{{\left\| {{\bf{r}}_m^{\left( i \right)} - {{\bf{r}}_b}} \right\|}_2}} \right)^T}\left( {{{\bf{r}}_m} - {\bf{r}}_m^{\left( i \right)}} \right)\\
		\;\;= {\left\| {{\bf{r}}_m^{\left( i \right)} - {{\bf{r}}_b}} \right\|_2} + \frac{1}{{{{\left\| {{\bf{r}}_m^{\left( i \right)} - {{\bf{r}}_b}} \right\|}_2}}}{\left( {{\bf{r}}_m^{\left( i \right)} - {{\bf{r}}_b}} \right)^T}\left( {{{\bf{r}}_m} - {\bf{r}}_m^{\left( i \right)}} \right)\\
		\;\;= \frac{1}{{{{\left\| {{\bf{r}}_m^{\left( i \right)} - {{\bf{r}}_b}} \right\|}_2}}}{\left( {{\bf{r}}_m^{\left( i \right)} - {{\bf{r}}_b}} \right)^T}\left( {{{\bf{r}}_m} - {{\bf{r}}_b}} \right).
	\end{array}
\end{equation}
\end{small}Then, (\ref{eqn24}e) is transformed into  

\begin{equation}\label{eqn37}
	\frac{1}{{{{\left\| {{\bf{r}}_m^{\left( i \right)} - {{\bf{r}}_b}} \right\|}_2}}}{\left( {{\bf{r}}_m^{\left( i \right)} - {{\bf{r}}_b}} \right)^T}\left( {{{\bf{r}}_m} - {{\bf{r}}_b}} \right) \ge {D_r},b \in {{\cal M}_{ - m}}.
\end{equation}Up to this point, $P$4 is able to be reformulated as

\begin{align}
	&	P4.1:{\rm{   }} \mathop {\max }\limits_{ {\bf{r}}_m}\;\;	\chi  \label{eqn38}\\
	&\;\;\;\;\;\;\;\;\;    {\rm{  s.t. }} \; (\ref{eqn24}d), (\ref{eqn29})-(\ref{eqn31}), (\ref{eqn37}). \nonumber
\end{align}Clearly, this constitutes a convex quadratically constrained program, and it can be efficiently solved using the existing tool CVX.

\subsection{Transmit Antenna Position Optimization}

Following the updates to parameters ${{\bf{u}}_r}$, $\left\{ {{{\bf{w}}_n}} \right\}_{n = 1}^N$, ${\bf{r}}$, and given ${{{\bf{t}}}_a}$ $\left( 1 \le a \ne n \le N \right)$, we proceed to optimize the position of the transmission antenna ${{{\bf{t}}}_n}$. Similarly to $P$4, we can reconstruct $P$1 to 
\begin{align}
	&	P5:{\rm{   }} \mathop {\max }\limits_{ {\bf{t}}_n, \alpha ,\beta }\;\;	\chi  \label{eqn39}\\
	&\;\;\;\;\;\;\;\;\;    {\rm{  s.t. }} \; \alpha  \ge \beta \chi , \tag{\ref{eqn39}{a}}\\
	&\;\;\;\;\;\;\;\;\;\;\;\;\;\;\;  f\left( {{{\bf{t}}_n}} \right) \ge \alpha , \tag{\ref{eqn39}{b}}\\
	&\;\;\;\;\;\;\;\;\;\;\;\;\;\;\;  g\left( {{{\bf{t}}_n}} \right) \le \beta , \tag{\ref{eqn39}{c}}\\
	&\;\;\;\;\;\;\;\;\;\;\;\;\;\;\;  \frac{{{{\left| {{\bf{h}}_k^H\left( {\bf{t}} \right){{\bf{w}}_k}} \right|}^2}}}{{\sum\nolimits_{{\widehat n} \ne k}^N {{{\left| {{\bf{h}}_k^H\left( {\bf{t}} \right){{\bf{w}}_{\widehat n}}} \right|}^2}}  + \delta _k^2}} \ge \gamma _k^{th}, k \in {\cal K}, \tag{\ref{eqn39}{d}}\\
	&\;\;\;\;\;\;\;\;\;\;\;\;\;\;\;  {{\bf{t}}_n} \in {{\cal C}_t} , \tag{\ref{eqn39}{e}}\\
	&\;\;\;\;\;\;\;\;\;\;\;\;\;\;\;  {\left\| {{{\bf{t}}_n} - {{\bf{t}}_a}} \right\|_2} \ge {D}, a \in {{\cal N}_{ - n}}. \tag{\ref{eqn39}{f}} 
\end{align}Similarly, we define ${{x_{n,\widehat n}}} = w_{\widehat n}^n\widetilde {\bf{g}}_{t,d}^H{\bf{f}}_{t,d}\left( {{{\bf{t}}_n}} \right)$, ${{y_{n,\widehat n,l}}} = w_{\widehat n}^n\widetilde {\bf{g}}_{t,{l}}^H{\bf{f}}_{t,{l}}\left( {{{\bf{t}}_n}} \right)$, and $\widetilde {\bf{g}}_{t,q}^H = {{\bf{u}}_r}{{\bf{h}}_{r,q}}\left( {\bf{r}} \right){\bf{g}}_{t,q}^H$, $q \in \left\{ {d,{l}} \right\}$. Then, $f\left( {{{\bf{r}}_m}} \right)$ and $g\left( {{{\bf{r}}_m}} \right)$ are given in (\ref{eqn40}) and (\ref{eqn41}), which are at the top of the next page. In this case, ${w_{\widehat n}^n}$ denotes the $n$-th element of ${{\bf{w}}_{\widehat n}}$. Furthermore, detailed derivations of the relevant terms for $\widetilde f\left( {{{\bf{t}}_m}} \right)$ and $\widetilde g\left( {{{\bf{t}}_m}} \right)$ in (\ref{eqn42}) and (\ref{eqn43}) are provided, which are available at the top of next page. Here, we have

\begin{figure*}[ht]
\begin{small}
	\begin{equation}\label{eqn40}
 f\left( {{{\bf{t}}_n}} \right) = \underbrace {{{\left| {{\alpha _d}} \right|}^2}\sum\limits_{\widehat n = 1}^N {\left( {{{\left| {{x_{n,\widehat n}}} \right|}^2} + 2\Re \left( {\sum\limits_{j = 1,j \ne n}^N {{x_{n,\widehat n}}x_{j,\widehat n}^*} } \right)}\!\!\! \right)} }_{\widetilde f\left( {{{\bf{t}}_n}} \right)} + {\left| {{\alpha _d}} \right|^2}\sum\limits_{\widehat n = 1}^N {\sum\limits_{j = 1,j \ne n}^N {\left( {{{\left| {{x_{j,\widehat n}}} \right|}^2} + 2\Re \left( {\sum\limits_{\widetilde j = j + 1,\widetilde j \ne n}^N { {{x_{\widetilde j,\widehat n}}x_{j,\widehat n}^*} } } \right)} \!\!\!\right)} }.
	\end{equation}
	\begin{equation}\label{eqn41}
		g\left( {{{\bf{t}}_n}} \right) = \underbrace {\sum\limits_{l = 1}^L {{{\left| {{\alpha _{{l}}}} \right|}^2}\!\sum\limits_{\widehat n = 1}^N \!\! {\left(\! {{{\left| {{y_{n,\widehat n,l}}} \right|}^2} \!\!+\!\! 2\Re \left( {\sum\limits_{j = 1,j \ne n}^N {{y_{n,\widehat n,l}}y_{j,\widehat n,l}^*} } \right)} \!\!\!\right)} } }_{\widetilde g\left( {{{\bf{t}}_m}} \right)} \!\!+\!\! \sum\limits_{l = 1}^L {{{\left| {{\alpha _{{l}}}} \right|}^2}\sum\limits_{\widehat n = 1}^N\!  {\sum\limits_{j = 1,j \ne n}^N {\left( {{{\left| {{y_{j,\widehat n,l}}} \right|}^2} \!\!+\!\! 2\Re \left( {\sum\limits_{\widetilde j = j + 1,\widetilde j \ne n}^N { {{y_{\widetilde j,\widehat n,l}}y_{j,\widehat n,l}^*} } } \right)}\!\!\! \right)} } }  \! +\!  \sigma _r^2.
	\end{equation}
\end{small}
	\begin{equation}\label{eqn42}
		\widetilde f\left( {{{\bf{t}}_n}} \right) = {\left| {{\alpha _d}} \right|^2}\sum\limits_{\hat n = 1}^N {\sum\limits_{j = 1}^{L_d^t} {\sum\limits_{p = 1}^{L_d^t} {\left[ {{{\left| {w_{\hat n}^n} \right|}^2}\left| {{{\left[ {{{\widetilde {\bf{G}}}_{d,\hat n}}} \right]}_{p,j}}} \right|\cos \lambda _{d,\hat n,j,p}^n + \sum\limits_{a = 1,a \ne m}^N {2\left| {w_{\hat n}^n} \right|\left| {w_{\hat n}^a} \right|\left| {{{\left[ {{{\widetilde {\bf{G}}}_{d,\hat n}}} \right]}_{p,j}}} \right|\cos \lambda _{d,\hat n,j,p}^a} } \right]} } } , 
	\end{equation}
		\begin{equation}\label{eqn43}
		\widetilde g\left( {{{\bf{t}}_n}} \right) = \sum\limits_{l = 1}^L {{{\left| {{\alpha _l}} \right|}^2}\sum\limits_{\hat n = 1}^N {\sum\limits_{j = 1}^{L_l^t} {\sum\limits_{p = 1}^{L_l^t} {\left[ {{{\left| {w_{\hat n}^n} \right|}^2}\left| {{{\left[ {{{\widetilde {\bf{G}}}_{l,\hat n}}} \right]}_{p,j}}} \right|\cos \lambda _{l,\hat n,j,p}^n + \sum\limits_{a = 1,a \ne m}^N {2\left| {w_{\hat n}^n} \right|\left| {w_{\hat n}^a} \right|\left| {{{\left[ {{{\widetilde {\bf{G}}}_{l,\hat n}}} \right]}_{p,j}}} \right|\cos \lambda _{l,\hat n,j,p}^a} } \right]} } } } .
	\end{equation}
\hrulefill
\end{figure*}
 
 \begin{small}
 	\begin{equation}\label{eqn102}
 		\lambda _{q,\hat n,j,p}^n = \frac{{2\pi }}{\lambda }\rho _{q,j}^t\left( {{{\bf{t}}_n}} \right) - \frac{{2\pi }}{\lambda }\rho _{q,p}^t\left( {{{\bf{t}}_n}} \right) + \angle {\left[ {{{\widetilde {\bf{G}}}_{q,\hat n}}} \right]_{p,j}},
 	\end{equation}
 \end{small}
 and
 \begin{small}
 	\begin{equation}\label{eqn103}
 		\lambda _{q,\hat n,j,p}^a = \frac{{2\pi }}{\lambda }\rho _{q,j}^t\left( {{{\bf{t}}_n}} \right) \!-\! \frac{{2\pi }}{\lambda }\rho _{q,p}^t\left( {{{\bf{t}}_a}} \right) \!+\! \angle {\left[ {{{\widetilde {\bf{G}}}_{q,\hat n}}} \right]_{p,j}} \!+\! \angle w_{\hat n}^n \!-\! \angle w_{\hat n}^a,
 	\end{equation}
 \end{small}where ${\widetilde {\bf{G}}_{q,{\widehat n}}} = {\widetilde {\bf{g}}_{q,{\widehat n}}}\widetilde {\bf{g}}_{q,{\widehat n}}^H$ and $q \in \left\{ {d,l} \right\}$. Similar to (\ref{eqn30}) and (\ref{eqn31}), the second-order Taylor expansions of functions $f\left( {{{\bf{t}}_n}} \right)$ and $g\left( {{{\bf{t}}_n}} \right)$ are denoted by $\overline f \left( {{{\bf{t}}_n}} \right)$ and $\overline g \left( {{{\bf{t}}_n}} \right)$, respectively. In addition, the computation method of parameters, including the gradient and the Hessian matrix, is detailed in the previous subsection and is not repeated here to maintain brevity. 

For (\ref{eqn39}d), we set ${x_{n,k}} = w_k^n{\bf{g}}_k^H{{\bf{f}}_k}\left( {{{\bf{t}}_n}} \right)$ and ${y_{n,k}} = w_{\widehat n}^n{\bf{g}}_k^H{{\bf{f}}_k}\left( {{{\bf{t}}_n}} \right)$. Then, we reconstruct the numerator and denominator as

	\begin{equation}\label{eqn44}
		\begin{array}{l}
			{f_k}\left( {{{\bf{t}}_n}} \right) = \underbrace {{{\left| {{x_{n,k}}} \right|}^2} + 2\Re \left( {\sum\limits_{j = 1,j \ne n}^N {{x_{n,k}} x_{j,k}^*} } \right)}_{{{\widetilde f}_k}\left( {{{\bf{t}}_n}} \right)}\\
			+ \sum\limits_{j = 1,j \ne n}^N {\left( {{{\left| {{x_{j,k}}} \right|}^2} + 2\Re \left( {\sum\limits_{\widetilde j = j + 1,\widetilde j \ne n}^N {\left( {{x_{\widetilde j,k}} x_{j,k}^*} \right)} } \right)} \!\!\! \right)}  
		\end{array}
	\end{equation}
and
	\begin{equation}\label{eqn45}
		\begin{array}{l}
			{g_k}\left( {{{\bf{t}}_n}} \right) = \underbrace {\sum\limits_{\hat n \ne k}^N {\left( {{{\left| {{y_{n,k}}} \right|}^2} + 2\Re \left( {\sum\limits_{j = 1,j \ne n}^N {{y_{n,k}} y_{j,k}^* } } \right)}\!\!\! \right)} }_{{{\widetilde g}_k}\left( {{{\bf{t}}_n}} \right)}\\
			\!\!\!+\! \sum\limits_{\hat n \ne k}^N \!\! {\left( \!{\sum\limits_{j = 1,j \ne n}^N\!\!\! {\left(\!\! {{{\left| {{y_{j,k}}} \right|}^2} \!+\! 2\Re \left( {\sum\limits_{\widetilde j = j + 1,\widetilde j \ne n}^N {\left( {{y_{\widetilde j,k}} y_{j,k}^*} \right)} } \right)}\!\!\! \right)} } \!\!\!\right)},
		\end{array}
	\end{equation}where the relevant terms ${{{\widetilde f}_k}\left( {{{\bf{t}}_n}} \right)}$ and ${{{\widetilde g}_k}\left( {{{\bf{t}}_n}} \right)}$ can similarly be derived using the methods previously, which are omitted here for brevity. Then, we employ the second-order Taylor expansion to transform (\ref{eqn39}d) into 

\begin{equation}\label{eqn46}
	{\overline f _k}\left( {{{\bf{t}}_n}} \right) \ge \gamma _k^{th}{\overline g _k}\left( {{{\bf{t}}_n}} \right), k \in {\cal K},
\end{equation}where ${\overline f _k}\left( {{{\bf{t}}_n}} \right)$ and ${\overline g _k}\left( {{{\bf{t}}_n}} \right)$ are the lower and upper bounds of $f_k\left( {{{\bf{t}}_n}} \right)$ and $g_k\left( {{{\bf{t}}_n}} \right)$ based on the second order Taylor expansion, respectively, and they can be derived using similar methods. Then, $P$5 is rewritten as 
\begin{align}
	&	P5.1:{\rm{   }} \mathop {\max }\limits_{ {\bf{t}}_n, \alpha ,\beta }\;\;	\chi  \label{eqn47}\\
	&\;\;\;\;\;\;\;\;\;    {\rm{  s.t. }} \; (\ref{eqn29}),(\ref{eqn39}e),(\ref{eqn46}), \nonumber\\
	&\;\;\;\;\;\;\;\;\;\;\;\;\;\;\;  \overline f \left( {{{\bf{t}}_n}} \right) \ge \alpha , \tag{\ref{eqn47}{a}}\\
	&\;\;\;\;\;\;\;\;\;\;\;\;\;\;\;  \overline g \left( {{{\bf{t}}_n}} \right) \le \beta , \tag{\ref{eqn47}{b}}\\
	&\;\;\;\;\;\;\;\;\;\;\;\;\;\;\;  	 \frac{{{{\left( {{\bf{t}}_n^{\left( i \right)} - {{\bf{t}}_a}} \right)}^T}\left( {{{\bf{t}}_n} - {{\bf{t}}_a}} \right)}}{{{{\left\| {{\bf{t}}_n^{\left( i \right)} - {{\bf{t}}_a}} \right\|}_2}}} \ge {D_t},a \in {{\cal N}_{ - n}}. \tag{\ref{eqn47}{c}} 
\end{align}Subsequently, the approach utilized for resolving $P$4.1 can be similarly employed to address $P$5.1.

\subsection{Overall Algorithm: Convergence and Complexity Analysis}

\begin{algorithm}[htbp]
	\caption{Alternating optimization algorithm for ${{\bf{u}}_r}$, $\left\{ {{{\bf{w}}_n}} \right\}_{n = 1}^N$, ${\bf{t}}$, and ${\bf{r}}$}
	\label{alg3}
	\begin{algorithmic}[1]
		
		\STATE Initialize the accuracy thresholds $\sigma $, number of iterations $iter{_{\max }}$, feasible points ${{\bf{u}}^0_r}$, $\left\{ {{{\bf{w}}^0_n}} \right\}_{n = 1}^N$, ${\bf{t}}^0$, and ${\bf{r}}^0$.
		
		\REPEAT
		
		\STATE 	Given $\left\{ {{{\bf{w}}^{iter}_n}} \right\}_{n = 1}^N$, ${\bf{t}}^{iter}$, and ${\bf{r}}^{iter}$, obtain  ${{\bf{u}}^{iter+1}_r}$ by (\ref{eqn20});
		
		\STATE 	Update ${{\bf{u}}^{iter+1}_r}$, given ${\bf{t}}^{iter}$ and ${\bf{r}}^{iter}$, obtain $\left\{ {{{\bf{w}}^{iter+1}_n}} \right\}_{n = 1}^N$ by solving $P$3; 		
		
		\FOR{$m=1$ to $M$}
		
		\STATE	Update ${{\bf{u}}^{iter+1}_r}$ and $\left\{ {{{\bf{w}}^{iter+1}_n}} \right\}_{n = 1}^N$, given ${\bf{t}}^{iter}$ and ${\bf{r}}_b^{iter}$,\;$\left( { 1 \le b \ne m \le N, } \right)$, obtained ${\bf{r}}_m^{iter+1}$ by solving $P$4.1; 		
		
		\ENDFOR
		
		\FOR{$n=1$ to $N$}
		
		\STATE	Update ${{\bf{u}}^{iter+1}_r}$, $\left\{ {{{\bf{w}}^{iter+1}_n}} \right\}_{n = 1}^N$, and ${\bf{r}}^{iter+1}$, given ${\bf{t}}_a^{iter}$\;$\left( { 1 \le a \ne n \le N, } \right)$, obtained ${\bf{t}}_n^{iter+1}$ by solving $P$5.1;		
		
		\ENDFOR
		
		\STATE $iter = iter + 1$;
		
		\UNTIL the increase in the objective value is below $\sigma $, or the number of iterations $ite{r_{\max }}$ is reached.
		
		\RETURN ${{\bf{u}}_r}$, $\left\{ {{{\bf{w}}_n}} \right\}_{n = 1}^N$, ${\bf{t}}$, and ${\bf{r}}$.
		
	\end{algorithmic}
\end{algorithm}
 
\subsubsection{Convergence}

The solution for $P$1 is summarized as {\bf Algorithm 1}, where the convergence is ensured by the presence of an upper power bound and the non-decreasing nature of each iteration. Moreover, the optimization order in our algorithm is designed to balance simplicity, efficiency, and solution quality by progressing from simpler subproblems (e.g., closed-form solutions for beamforming) to more complex ones (e.g., MA position optimization). This approach ensures convergence.

\subsubsection{Complexity Analysis}

Subsequently, the complexity of our proposed scheme is mainly due to the implementation of {\bf Algorithm 1}. Obviously, it depends on the complexity of each iteration and the number of iterations. First, we focus on the computational complexity of $\left\{ {{{\bf{u}}_r},\left\{ {{{\bf{w}}_n}} \right\}_{n = 1}^N,{\bf{t}},{\bf{r}}} \right\}$ in each iteration. The update process for ${{\bf{u}}_r}$ primarily depends on computing the matrix inverse, characterized by ${\mathcal O}(M^3)$. For updating $\left\{ {{{\bf{w}}_n}} \right\}_{n = 1}^N$, according to \cite{r9}, the number of iterations is $\sqrt {{\beta _w}}  = \sqrt {{N^2} + K + 3} $ and the complexity of each iteration is ${{\rm{C}}_w} = {n_w}\left( {{N^4} + K + 3} \right) + n_w^2\left( {{N^3} + K + 3} \right) + n_w^3$ where ${n_w} = {\cal O}({N^3})$, and hence, the complexity of $P$3.1 is $\sqrt {{\beta _w}} {{\rm{C}}_w}$. For updating ${{{\bf{r}}}}$, we adopt a sequential optimization for each antenna position. Therefore, the complexity is mainly due to solving the subproblem $P$4.1 and the number of Rx antennas $M$. For $P$4.1, the number of iterations and the complexity of each iteration are $\sqrt {{\beta _{{{\bf{r}}_m}}}}  = \sqrt {2M + 7}$ and ${{\rm{C}}_{{{\bf{r}}_m}}} = {n_{{{\bf{r}}_m}}}\left( {M + 13} \right) + n_{{{\bf{r}}_m}}^2\left( {M + 9} \right) + n_{{{\bf{r}}_m}}^3$, respectively, in which ${n_{{{\bf{r}}_m}}} = {\cal O}(1)$. Thus, the complexity of $P$4.1 is $\sqrt {{\beta _{{{{\bf{r}}_m}}}}} {{\rm{C}}_{{{{\bf{r}}_m}}}}$, and the complexity of updating ${{{\bf{r}}}}$ is $M\sqrt {{\beta _{{{{\bf{r}}_m}}}}} {{\rm{C}}_{{{{\bf{r}}_m}}}}$. Similarly, for updating ${{{\bf{t}}}}$, the sequential optimization for each antenna position is also adopted. Thus, the complexity is mainly due to solving the subproblem $P$5.1 and the number of Tx antennas $N$. The complexity of $P$5.1 is $\sqrt {{\beta _{{{\bf{t}}_n}}}} {{\rm{C}}_{{{\bf{t}}_n}}}$, where $\sqrt {{\beta _{{{\bf{t}}_n}}}}  = \sqrt {2\left( {N + K} \right) + 7} $, ${{\rm{C}}_{{{\bf{t}}_n}}} = {n_{{{\bf{t}}_n}}}\left( {M + 13} \right) + n_{{{\bf{t}}_n}}^2\left( {M + 9} \right) + n_{{{\bf{t}}_n}}^3$, and ${n_{{{\bf{t}}_n}}} = {\cal O}(1)$. Thus, the complexity of updating ${{{\bf{t}}}}$ is $N\sqrt {{\beta _{{{\bf{t}}_n}}}} {{\rm{C}}_{{{\bf{t}}_n}}}$. Furthermore, the convergence of the algorithm ensures that the number of iterations remains within an acceptable range, with the computational complexity predominantly concentrated in solving the aforementioned subproblems.

\section{SIMULATION RESULTS}

In this section, we present simulation results to evaluate the performance of the proposed MA-assisted ISAC system. First, we set $K=2$ and $L=2$. The simulation considers a 3D coordinate system where the Tx is positioned at $(0,0,5)$ and the Rx at $(0,20,5)$. The target and clutters are strategically placed at $(0,10,5)$, $(10,10,5)$, and $(-10,10,5)$, respectively, while user locations are set at $(15,10,0)$ and $(-15,10,0)$. We consider the geometric channel model and set the number of transmit paths and receive paths to $L_k^t = L_q^t = L_q^r = 4$. Then, the PRV parameters of the line of sight (LoS) and non-LoS (NLoS) paths for ${{\bf{g}}_{k}}$ are modeled as ${g_1} \sim \mathcal{CN} \left( {0,{\beta _0}{{  {d_k/} }^{ - {\alpha _k}}}\kappa /\left( {\kappa  + 1} \right)} \right)$ and ${g_l}\ \sim \mathcal{CN} \left( {0,{\beta _0}{{ {d_k}  }^{ - {\alpha _k}}}/\left( {\left( {\kappa  + 1} \right)\left( {L_k^t - 1} \right)} \right)} \right)$, $l = 2,3, \ldots ,L_k^t$, where ${{\beta _0}} = -30$ dBm denotes the path loss per unit distance, ${d_k}$ is the distance from Tx to user $k$, ${{\alpha _k}}$ is the path loss exponent, and $\kappa $ is the ratio of the power of the LoS path to the NLoS path. Similarly, the PRVs for the remaining channels are configured appropriately. The path loss exponents of BS to the user, target, and clutters are 2.5, 2.2, and 2.3, respectively. The communication SINR thresholds are set to $\gamma _k^{th} = \gamma ^{th}  = 0$dB. 


\emph{Tag/Benchmark:} We designate the proposed {\bf Algorithm 1} as ``Proposed". To evaluate its performance, we contrast this scheme with three benchmark strategies. 1) {\bf{Receive--MA}}: The Tx comprises a PA consisting of $N$ FPAs, with adjacent antennas spaced by ${\frac{\lambda }{2}}$. The Rx employs a PA consisting of $M$ MAs for signal reception. 2) {\bf{Transmit--MA}}: The Tx is a PA comprising $N$ MAs, and the Rx is a PA consisting of $M$ FPAs with adjacent antennas spaced by ${\frac{\lambda }{2}}$. 3) {\bf{FPA}}: The Tx and Rx each employ a PA consisting of $N$ and $M$ FPAs, respectively, with antenna spacing of ${\frac{\lambda }{2}}$.

\subsection{Convergence Performance of Proposed Algorithms}

\begin{figure}
	\centering
	\includegraphics[width=0.46\textwidth,angle=0]{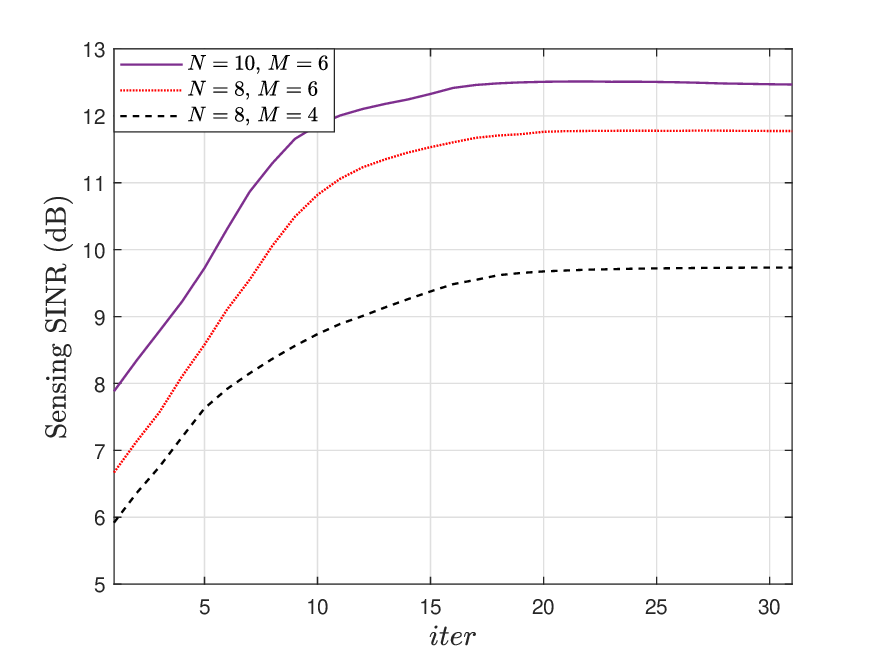}
	\centering
	\caption{Convergence performance of the proposed algorithms.}
	\label{Fig_2}
\end{figure}

In Fig. \ref{Fig_2}, we present a plot of the received SINR of the radar signal versus the number of iterations, illustrating the convergence performance of {\bf Algorithm 1}. The graph demonstrates that the performance of configurations with varying numbers of antennas generally achieves convergence after 20 iterations, with the antenna count having a minor impact on the convergence rate. Furthermore, the increase in sensing SINR becomes negligible after approximately 20 iterations, confirming the rapid convergence and effectiveness of the algorithm

\subsection{Impact of Transmit Power}

\begin{figure}
	\centering
	\includegraphics[width=0.46\textwidth,angle=0]{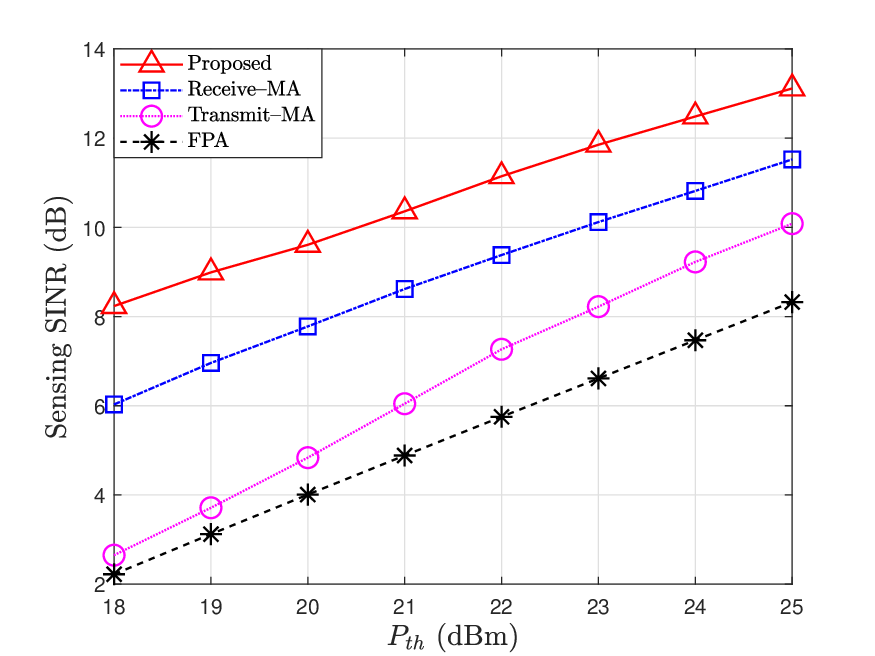}
	\centering
	\caption{Sensing SINR versus transmit power $P_{th}$. ($N=10$, $M=6$)}
	\label{Fig_3}
\end{figure}

In Fig. \ref{Fig_3}, we plot the sensing SINR versus the transmit power of Tx obtained by different schemes with Tx antennas $N=10$ and Rx antennas $M=6$. Initially, it is evident that the sensing SINR consistently increases with transmit power across all schemes. Significantly, our proposed scheme demonstrates superior performance, e.g., the Proposed scheme improves the performance by about $57.54\%$ over the FPA scheme at $P_{th}=25$ dBm. Notably, substantial performance improvements are also realized with MA configurations alone on either the Tx or Rx side, with the Receive-MA and Transmit-MA schemes yielding performance enhancements of approximately $38.43\%$ and $21.12\%$ over the FPA scheme, respectively. Additionally, it is observed that the performance advantage of Transmit-MA over FPA is less at lower power levels. This is due to the fact that at low power, the system relies more on the base gain of the antenna and the characteristics of the channel itself, and thus the Transmit-MA scheme performs similarly to the FPA.

\subsection{Impact of Number of Tx Antennas}

\begin{figure}
	\centering
	\includegraphics[width=0.46\textwidth,angle=0]{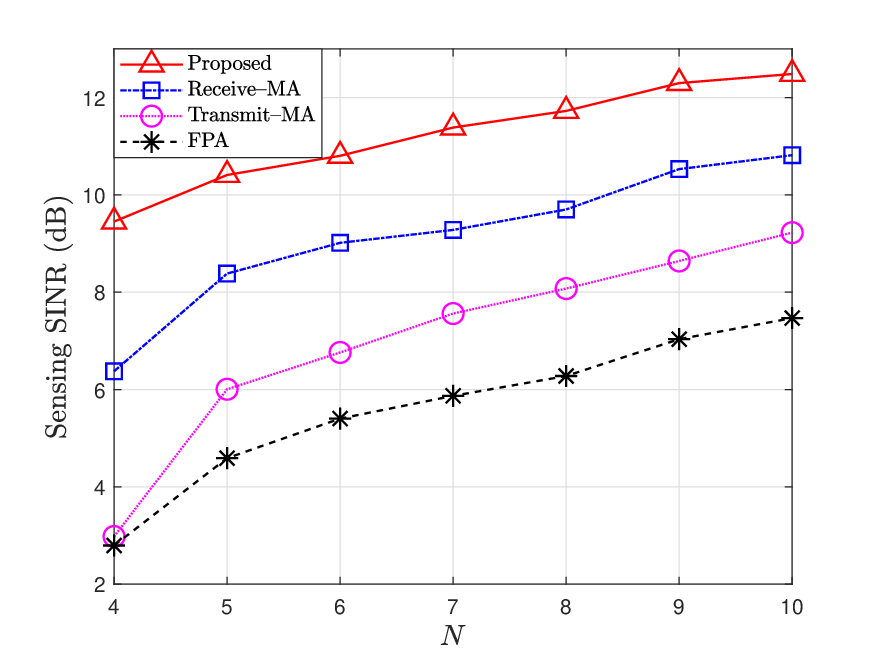}
	\centering
	\caption{Sensing SINR versus the number of Tx antennas $N$. ($P_s = 24$ dBm, $M=6$)}
	\label{Fig_4}
\end{figure}

Fig. \ref{Fig_4} illustrates the variation in the sensing SINR across various schemes as the number of Tx antennas ($N$) increases, with a transmit power of $P_s = 24$ dBm and Rx antennas $M=6$. The Proposed scheme consistently outperforms all others, with the sensing SINR progressively improving from approximately $9.45$ dB at $N=4$ to about $12.49$ dB at $N=10$. This enhancement is primarily due to the optimization strategy, which facilitates a more efficient exploitation of spatial resources in scenarios with multiple MAs. Receive-MA scheme significantly outperforms the Transmit-MA and FPA schemes in all cases, and the SINR increases from about $6.37$ dB to $10.82$ dB as the number of antennas increases. The relative performance growth is due to better suppression of interference from clutters by the MAs array at the Rx. For the Transmit-MA scheme, it also outperforms the FPA scheme, with an increase in sensing SINR from about $2.98$ dB to $9.22$ dB. The gain of this scheme over FPA comes from the MA optimization at Tx, which improves the directionality of the transmitter and the signal coverage, but it is slightly inferior to the Receive-MA scheme. Ultimately, the lower performance of the FPA scheme attributes to its limited flexibility and dynamic adaptability, characteristic of MA systems.

\subsection{Impact of Number of Rx Antennas}

\begin{figure}
	\centering
	\includegraphics[width=0.46\textwidth,angle=0]{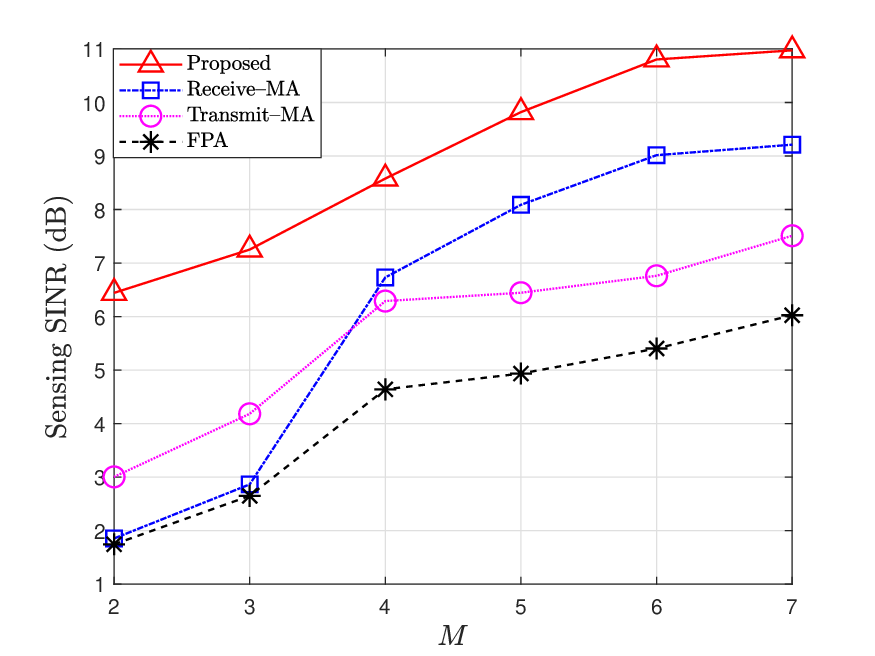}
	\centering
	\caption{Sensing SINR versus the number of Rx antennas $M$. ($P_s = 24$ dBm, $N=6$)}
	\label{Fig_5}
\end{figure}

In Fig. \ref{Fig_5}, the variation of the sensing SINR is shown for the different schemes when the number of receive antennas $M$ is increased, with a transmit power of $P_s = 24$ dBm and the number of Tx antennas $N=6$. The Proposed scheme outperforms all others, with the sensing SINR consistently increasing from approximately $6.44$ dB to $10.97$ dB as $M$ increases. By dynamically adjusting the transmit and receive strategies, this scheme significantly enhances signal gain and interference suppression. The Receive-MA scheme exhibits performance gains comparable to the FPA scheme when the number of $M < 4$. As $M$ increases, the sensing SINR consistently rises from approximately $1.85$ dB to $9.21$ dB, ultimately surpassing the FPA scheme significantly. Moreover, the Transmit-MA scheme is noticeably less effective than the Receive-MA scheme when $M > 5$, and it consistently outperforms the FPA scheme. This is due to the limited interference suppression capability of the FPA antenna at the Rx, but the signal strength and coverage can be enhanced by MAs at the Tx.

\subsection{Impact of Communication SINR}
 
\begin{figure}
	\centering
	\includegraphics[width=0.5\textwidth,angle=0]{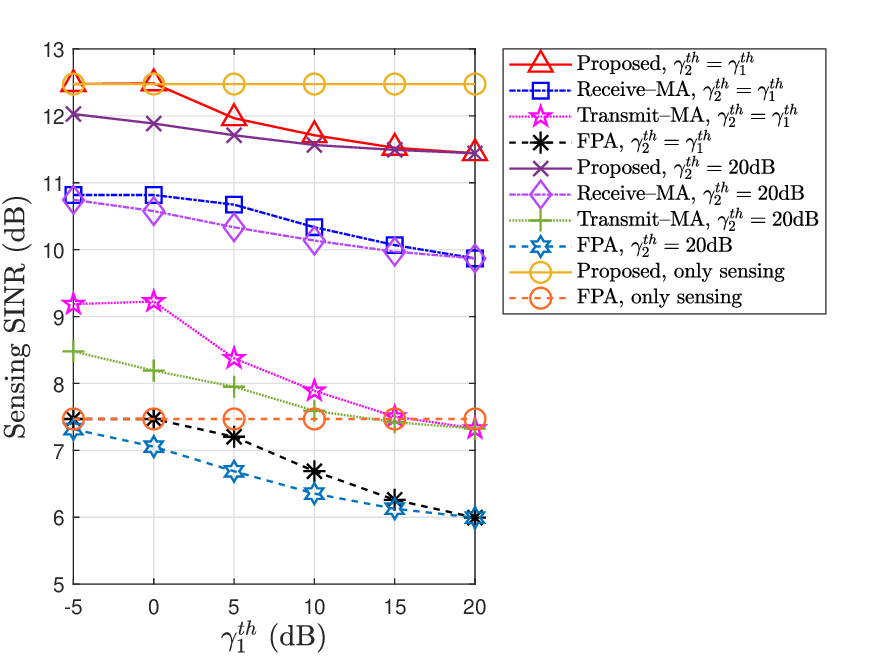}
	\centering
	\caption{Sensing SINR versus the communication SINR $\gamma_1^{th}$. ($P_s = 24$ dBm, $N=10$, $M=6$)}
	\label{Fig_6}
\end{figure}

Fig. \ref{Fig_6} illustrates two cases. One is the variation of the sensing SINR of the different schemes when the communication SINR threshold $\gamma_1^{th} = \gamma_2^{th}$ is varied, with the simulation conditions set to $P_s = 24$ dBm, $N=10$, and $M=6$. The other only $\gamma_1^{th}$ varies ($\gamma_2^{th}=20$ dB) and the changes of sensing SINR under different schemes.

For $\gamma_1^{th} = \gamma_2^{th}$, the Proposed scheme consistently delivers better performance across all communication SINR thresholds. The sensing SINR exhibits a minor decline from $12.48$ dB to $11.44$ dB as $\gamma_1^{th}$ increases, but the overall performance remains remarkably stable. This scheme exemplifies the robustness of effectively balancing resource allocation between communication and sensing, ensuring sustained perceptual quality even under increased communication loads, particularly at higher $\gamma_1^{th}$ values. The Receive-MA scheme experiences a modest reduction in perceived SINR, declining from $10.82$ dB to $9.87$ dB as $\gamma_1^{th}$ increases. Despite this slight performance drop, it consistently surpasses both the Transmit-MA and FPA schemes. However, the Transmit-MA scheme exhibits a more pronounced decrease in sensing SINR, from $9.19$ dB to $7.32$ dB as $\gamma_1^{th}$ escalates from $-5$ dB to $20$ dB. Furthermore, the performance trajectory of the FPA scheme illustrates the inherent limitations of traditional fixed antenna configurations in ISAC applications.

 For $\gamma_2^{th}=20$dB and $\gamma _1^{th}$ varies, we find that the total communication demand at the beginning is greater than that in $\gamma_1^{th} = \gamma_2^{th}$, so their sensing SINR decreases faster as $\gamma_1^{th}$ increases. The performance can overlap until their communication demands are the same. In addition, the performance trends of different MA schemes are consistent with he case of  $\gamma_1^{th} = \gamma_2^{th}$. In summary, the overall results, including those from other evaluations, show that the MA technology at the receiver end proves to be highly effective in enhancing sensing performance. By considering heterogeneous SINR thresholds for two users, we demonstrate the robustness and flexibility of the proposed scheme in balancing sensing and communication demands under realistic conditions.

In addition, we can observe that the ISAC scheme performs comparably to the sensing-only system at lower communication SINR thresholds. This is because the dual-purpose beamforming efficiently balances sensing and communication requirements without significantly compromising sensing performance. As $\gamma _1^{th}$ increases, the trade-off between communication and sensing objectives becomes more pronounced. This results in a slight degradation of sensing SINR for the proposed ISAC scheme. However, the flexibility provided by MAs mitigates the performance degradation caused by increased communication SINR thresholds. For example: at $\gamma _1^{th} = \gamma _2^{th} = 20$ dB, the proposed MA-assisted ISAC scheme experiences $1$ dB reduction in sensing SINR, compared to $1.5$ dB reduction for the FPA scheme. This highlights the benefits of the additional DoFs provided by MAs in balancing sensing and communication objectives.

\subsection{Impact of Size of Moveable Area}

\begin{figure}
	\centering
	\includegraphics[width=0.46\textwidth,angle=0]{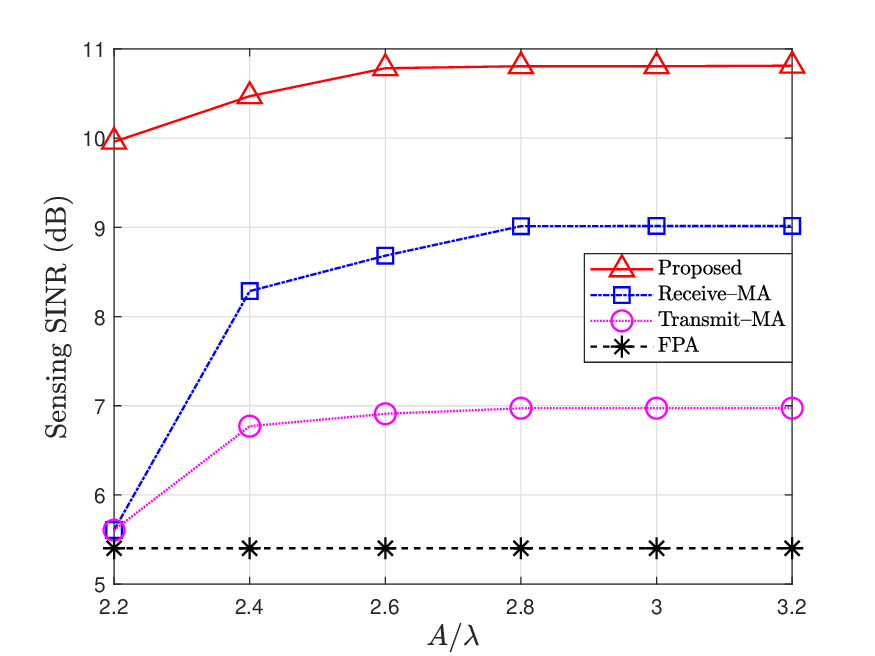}
	\centering
	\caption{Sensing SINR versus the size of the moveable area. ($P_s = 24$ dBm, $N=10$, $M=6$)}
	\label{Fig_7}
\end{figure}

In Fig. \ref{Fig_7}, the effect of the mobile range $A/\lambda$ of the movable antenna on the sensing SINR is demonstrated with the simulation parameters set to $P_s = 24$ dBm, $N=10$, and $M=6$. The Proposed scheme sustains high gains even at the limited movement range of $A/\lambda = 2.2$, significantly surpassing configurations where MAs are implemented solely at either the Tx or Rx, achieving a performance gain of approximately $77.67\%$. This gain is attributed to the synergistic effect of the dynamic positioning of MAs on both sides, and the total antenna space is greater than that of a single-sided deployment. In addition, it was observed that when $A/\lambda  = 2.6$, the gain of the proposed scheme reaches its maximum, indicating that the spatial requirements of the antenna are not unbounded. This finding supports preserving the miniaturization trend of antennas when deploying MA in future applications. Both the Receive-MA and Transmit-MA schemes reach a plateau when $A/\lambda > 2.8$, suggesting that the maximum sensing SINR can be achieved within the constrained transmit and/or receive area, and their performance remains inferior to that of the proposed scheme. This result demonstrates that, beyond enabling antenna miniaturization on both sides, dual-sided deployment offers significantly superior performance compared to single-sided configurations and FPA.

\subsection{Beampattern and Channel Characteristics}

\begin{figure}
	\centering
	\includegraphics[width=0.45\textwidth,angle=0]{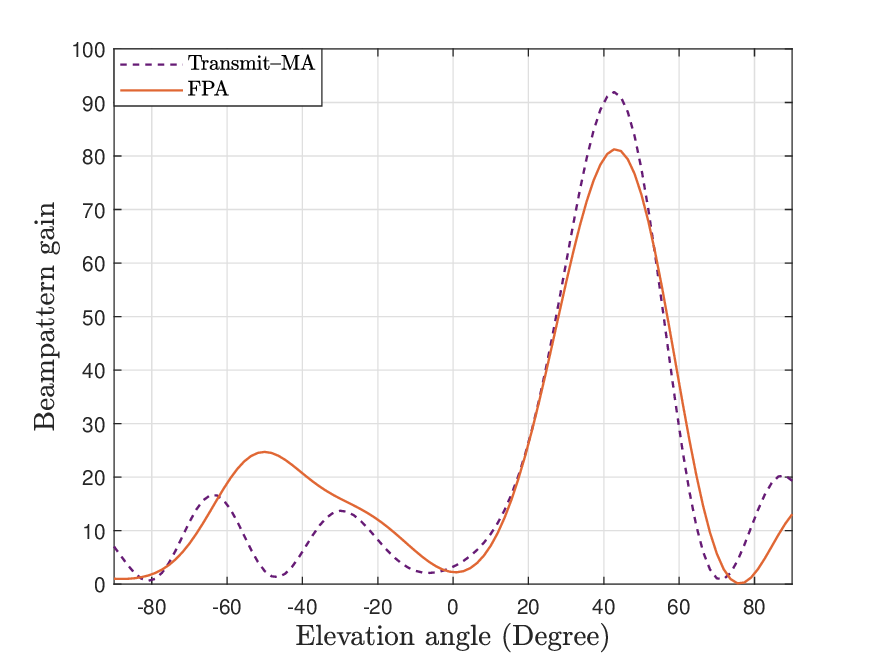}
	\centering
	\caption{Beampattern of Tx. ($N=10$)} 
	\label{Fig_8}
\end{figure}

Fig. \ref{Fig_8} shows the beampattern of Tx, which mainly shows the beam gain distribution of the two schemes at different elevation angles, and the target elevation angle is set to be ${45^ \circ }$. The beampattern gain of the Transmit-MA scheme has an enhancement compared with that of the FPA scheme in the direction of the angle of interest. In addition, the Transmit-MA scheme performs better than the FPA scheme in controlling the gain in the uninterested direction. Meanwhile, the main lobe of the Transmit-MA scheme exhibits a narrower profile, indicating that beam orientation can be  refined through dynamic adjustments of the MAs.

\begin{figure}
	\centering
	\includegraphics[width=0.45\textwidth,angle=0]{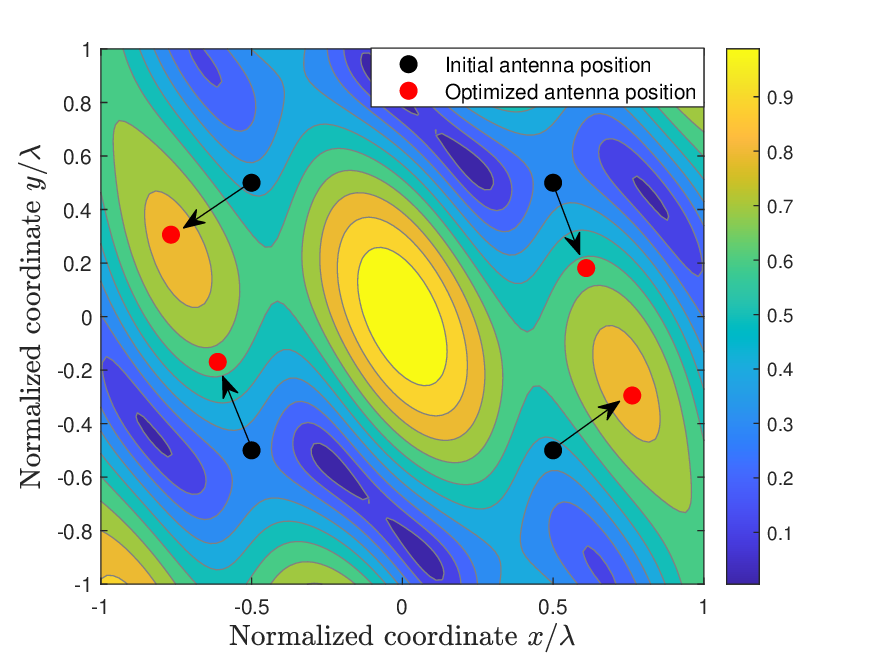}
	\centering
	\caption{Example of channel power gain in Rx area ${{\cal C}_r}$. ($M=4$)}
	\label{Fig_9}
\end{figure}

The primary driver of performance enhancement through the deployment of MAs is their capability to reconfigure the channel response. For this reason, we show in Fig. \ref{Fig_9} the relation between the channel gain of the target-Rx link and the positions of the MAs. Evidently, the optimization algorithm strategically moves the antenna positions to regions with higher power gain.
The optimization of antenna positions should consider both the enhancement of the gain along the primary signal path and the suppression of interfering signals in undesired directions. This allows, unlike the single antenna optimized to the location of maximum gain in \cite{r10}, a multi-antenna receiving array at a specific location to form the desired beampattern. Thus, this flexible DoFs offers MAs the potential to achieve significant performance compared to FPAs.

\section{CONCLUSION}

In this paper, we have investigated MA-assisted ISAC systems, where MAs are strategically positioned at both the Tx and Rx of the BS to facilitate dual functionalities of sensing and communication. To improve the ISAC performance, we have formulated an optimization problem to maximize the sensing SINR by jointly optimizing the received vector, the transmitted matrix, and the antenna positions subject to the minimum communication SINR requirement. Given the highly non-convex nature of the problem, we have developed an AO-based algorithm employing techniques such as the Charnes-Cooper transformation, second-order Taylor expansion, and SCA. Simulation results demonstrate that the proposed MA-assisted ISAC system offers substantial benefits over baseline schemes, and the performance gain over FPA can also be obtained by deploying MA only at the Rx or Tx. This shows that MAs provide additional spatial degrees of freedom, enabling the system to dynamically reconfigure antenna positions to maximize beamforming gains, suppress interference, and improve overall performance. In particular, the proposed scheme demonstrates the capability to match the performance of the FPA scheme while utilizing fewer resources, e.g., the sensing SINR achieved by the proposed scheme at a transmit power of 18 dBm is comparable to that obtained by the FPA scheme at 25 dBm. The proposed system demonstrates a performance enhancement exceeding $57.54\%$ over FPA when operating at equivalent power levels. Moreover, the ability to reposition antennas dynamically enables the system to adapt to changing environmental conditions, ensuring stable performance. In the future, to ensure computational efficiency and maintain high performance, advanced technologies such as distributed optimization, cooperative beamforming, and layered antenna control can be investigated to address the challenges of large-scale ISAC systems. Furthermore, exploring the synergies between MA and other advanced hardware integration technologies represents a promising avenue for future research.

\begin{appendices}
	\section{The proof process of Theorem 1}
	\emph{Proof:} Firstly, we can obtain the Lagrangian function for $P$3.2 in (\ref{eqn57}) at the top of next page, in which $\zeta \ge 0$, $\left\{ {{{\bf{Z}}_n}} \right\}_{n = 1}^N \succeq 0$, $\hbar \ge 0$, $\psi \ge 0$, $\left\{ {{\mathchar'26\mkern-10mu\lambda _k}} \right\}_{k = 1}^K \ge 0 $. Then, by taking the partial derivatives of the Lagrangian function in (\ref{eqn57}) with respect to ${{\bf{X}}_k}$, $1 \le k \le K$, the KKT condition can be derived as

	\begin{figure*}[ht]
		\begin{footnotesize}
		\begin{equation}\label{eqn57} 
		\begin{array}{l}
			{\cal L}  \left( {\zeta ,\left\{ {{{\bf{Z}}_n}} \right\}_{n = 1}^N,\hbar ,\psi ,\left\{ {{\mathchar'26\mkern-10mu\lambda _k}} \right\}_{k = 1}^K} \right) \! = \! \sum\nolimits_{n = 1}^N {{\rm{tr}}\left( {{{\bf{X}}_n}} \right)}  \!-\! \zeta \left( {{{\left| {{\alpha _d}} \right|}^2}\sum\nolimits_{n = 1}^N {{\rm{tr}}\left( {{{\widetilde {\bf{H}}}_d}{{\bf{X}}_n}} \right) \!-\! {\gamma ^\star}} } \right) \!+\! \psi \left( {\sum\nolimits_{l = 1}^L {{{\left| {{\alpha _{{l}}}} \right|}^2}\sum\nolimits_{n = 1}^N {{\rm{tr}}\left( {{{\widetilde {\bf{H}}}_{{l}}}{{\bf{X}}_n}} \right)} }  \!+\! \sigma _r^2\ell  \!-\! 1} \right)  \\
			\;\;\;\;\;\;\;\;\;\;\;\;\;\;\;\;\;\;\;\;\;\;\;\;\;\;\;\;\;\;\;\;\;\;\;\;\;\;\;\;\;\;\;\;\;\;\;\;\;\;\;\;\; - \sum\nolimits_{n = 1}^N {{{\bf{Z}}_n}{{\bf{X}}_n}} - \hbar \ell  - \sum\nolimits_{k = 1}^K {{\mathchar'26\mkern-10mu\lambda _k}\left( {{\rm{tr}}\left( {{{\widehat {\bf{H}}}_k}{{\bf{X}}_k}} \right) - \gamma _k^{th}\sum\nolimits_{n \ne k}^N {{\rm{tr}}\left( {{{\widehat {\bf{H}}}_k}{{\bf{X}}_n}} \right)}  - \ell \gamma _k^{th}\delta _k^2  } \right)} .
		\end{array}
		\end{equation}
		\hrulefill
		\end{footnotesize}
	\end{figure*}

	\begin{equation}\label{eqn58}
		{\bf{A}} - \zeta {\widetilde {\bf{H}}_d} - {{\bf{Z}}_k} - {\mathchar'26\mkern-10mu\lambda _k}{\widehat {\bf{H}}_k} = 0,
	\end{equation}
	\begin{equation}\label{eqn59}
		\;\;\;\;\;\;\;\;\;\;\;\;\;\;\;\;\;\;\;\;\;\;\;\;\;\;\;\;     {{\bf{Z}}_k}{{\bf{X}}_k} = 0,
	\end{equation}
	\begin{equation}\label{eqn60}
		\;\;\;\;\;\;\;\;\;\;\;\;\;\;\;\;\;\;\;\;\;\;\;\;\;\;\;\;\;\;\;\;\;    {{\bf{X}}_k} \succeq 0,
	\end{equation}where ${\bf{A}} = {{\bf{I}}_N} + \psi \sum\nolimits_{l = 1}^L {{{\left| {{\alpha _{{l}}}}  \right|}^2}{{\widetilde {\bf{H}}}_{{l}}}}  + \sum\nolimits_{j \ne k}^K {{\mathchar'26\mkern-10mu\lambda _j}\gamma _j^{th}{{\widehat {\bf{H}}}_j}}  $. Evidently, we find that ${\bf{A}} \succ 0$ holds due to $\psi, {{\mathchar'26\mkern-10mu\lambda _j}}   \ge 0$. Then, we define ${\bf{Z}}_k = {\bf{A}} - \zeta {\widetilde {\bf{H}}_d}  -{\mathchar'26\mkern-10mu\lambda _k}{\widehat {\bf{H}}_k}$. In addition, we find ${\rm{rank}}\left( {{{\widetilde {\bf{H}}}_d}} \right) = 1$ and ${\rm{rank}}\left( {{{\widehat {\bf{H}}}_k}} \right) = 1$ due to their definitions. Due to the randomness of the channel, the maximum eigenvalues of ${{{\widetilde {\bf{H}}}_d}}$ and ${{{\widehat {\bf{H}}}_k}}$ are different\cite{r47}, and in order to ensure that ${\bf{Z}}_k \succeq 0$, we can conclude that  
	
	\begin{equation}\label{eqn61}
		N - 1 \le {\rm{rank}}\left( {\bf{Z}}_k \right) \le N.
	\end{equation}Meanwhile, using Sylvester inequality, we can obtain

	\begin{equation}\label{eqn62}
	   {\rm{rank}}\left( {\bf{Z}}_k \right) + {\rm{rank}}\left( {{{\bf{X}}_k}} \right) \le {\rm{rank}}\left( {{\bf{Z}}_k{{\bf{X}}_k}} \right) + N.
    \end{equation}Clearly, we can find that ${\rm{rank}}\left( {{\bf{Z}}_k{{\bf{X}}_k}} \right) = 0$, then ${\rm{rank}}\left( {\bf{Z}}_k \right) + {\rm{rank}}\left( {{{\bf{X}}_k}} \right) \le  N$, and again by ${\rm{rank}}\left( {\bf{Z}}_k \right) \ge N - 1$, we can conclude that $ {\rm{rank}}\left( {{{\bf{X}}_k}} \right) \le  1$. By discarding the trivial solution $ {\rm{rank}}\left( {{{\bf{X}}_k}} \right) =  0$, it is proved $ {\rm{rank}}\left( {{{\bf{X}}_k}} \right) =  1$. Thus, ${rank\left( {{{\bf{X}}_k}} \right)} = 1$, $1 \le k \le K$ is proved.
    
    Next, we prove the rank condition of ${{\bf{X}}_q}, {K +1 \le q \le N}  $. Similarly, by solving for the partial derivatives of the Lagrangian function with respect to ${{\bf{X}}_q}$ in (\ref{eqn57}), the KKT condition can be obtained as

	\begin{equation}\label{eqn65}
		{\bf{B}} - \zeta {\widetilde {\bf{H}}_d} - {{\bf{Z}}_q} = 0,
	\end{equation}
	\begin{equation}\label{eqn66}
		\;\;\;\;\;\;\;\;\;\;\;\;\;\;\;      {{\bf{Z}}_q}{{\bf{X}}_q} = 0,
	\end{equation}
	\begin{equation}\label{eqn67}
		\;\;\;\;\;\;\;\;\;\;\;\;\;\;\;\;\;\;\;\;      {{\bf{X}}_q} \succeq 0,
	\end{equation}where ${\bf{B}} = {{\bf{I}}_N} + \psi \sum\nolimits_{l = 1}^L {{{\left| {{\alpha _{{l}}}} \right|}^2}{{\widetilde {\bf{H}}}_{{l}}}}  + \sum\nolimits_{k = 1}^K {{\mathchar'26\mkern-10mu\lambda _k}\gamma _k^{th}{{\widehat {\bf{H}}}_k}} $. Likewise, ${\bf{B}} \succ 0$ holds due to $\psi, {{\mathchar'26\mkern-10mu\lambda _k}}   \ge 0$. By post-multiplying both sides of (\ref{eqn65}) with ${{{\bf{X}}_k}}$ and combining with (\ref{eqn66}), we obtain ${\bf{B}}{{\bf{X}}_q} = \zeta {\widetilde {\bf{H}}_t}{{\bf{X}}_q}$. Based on the above, we find that
	
	\begin{equation}\label{eqn68}
		{\rm{rank}}\left( {{\bf{B}}{{\bf{X}}_k}} \right) = {\rm{rank}}\left( {{{\bf{X}}_k}} \right) = {\rm{rank}}\left( {\zeta {{\widetilde {\bf{H}}}_d}{{\bf{X}}_q}} \right) \le 1.
	\end{equation}
	
	After discarding the trivial case ${\rm{rank}}\left( {{{\bf{X}}_q}} \right) = 0$, ${\rm{rank}}\left( {{{\bf{X}}_q}} \right) = 1,K + 1 \le q \le N$ is proved. $\hfill\blacksquare$

\section{The gradient and Hessian matrices.}
	The gradients $\nabla f\left( {{{\bf{r}}_m}} \right) = {\left[ {\frac{{\partial f\left( {{{\bf{r}}_m}} \right)}}{{\partial {x_m}}}{\rm{ }}\frac{{\partial f\left( {{{\bf{r}}_m}} \right)}}{{\partial {y_m}}}} \right]^T}$ and $\nabla g\left( {{{\bf{r}}_m}} \right) = {\left[ {\frac{{\partial g\left( {{{\bf{r}}_m}} \right)}}{{\partial {x_m}}}{\rm{ }}\frac{{\partial g\left( {{{\bf{r}}_m}} \right)}}{{\partial {y_m}}}} \right]^T}$ are specified in (\ref{eqn69})-(\ref{eqn72}). Accordingly, we give the following symbolic settings for simplified expressions, i.e.,
		
\begin{equation}\label{eqn104}
			\dot \theta _{j,p}^q = \left( {\sin \phi _{q,p}^r\cos \theta _{q,p}^r - \sin \phi _{q,j}^r\cos \theta _{q,j}^r} \right),
\end{equation}
and
\begin{equation}\label{eqn105}
			\ddot \theta _{j,p}^q = \left( {\cos \phi _{q,p}^r - \cos \phi _{q,j}^r} \right),
\end{equation}where $q \in \left\{ {d,l} \right\}$. Furthermore, the components of the Hessian matrices ${\nabla ^2}g\left( {{{\bf{r}}_m}} \right) = \left[ \begin{array}{l}
		\frac{{{\partial ^2}g\left( {{{\bf{r}}_m}} \right)}}{{\partial {x_m}\partial {x_m}}}{\rm{ }}\frac{{{\partial ^2}g\left( {{{\bf{r}}_m}} \right)}}{{\partial {x_m}\partial {y_m}}}\\
		\frac{{{\partial ^2}g\left( {{{\bf{r}}_m}} \right)}}{{\partial {y_m}\partial {x_m}}}{\rm{ }}\frac{{{\partial ^2}g\left( {{{\bf{r}}_m}} \right)}}{{\partial {y_m}\partial {y_m}}}
	\end{array} \right]$ and ${\nabla ^2}f\left( {{{\bf{r}}_m}} \right) = \left[ \begin{array}{l}
	\frac{{{\partial ^2}f\left( {{{\bf{r}}_m}} \right)}}{{\partial {x_m}\partial {x_m}}}{\rm{ }}\frac{{{\partial ^2}f\left( {{{\bf{r}}_m}} \right)}}{{\partial {x_m}\partial {y_m}}}\\
	\frac{{{\partial ^2}f\left( {{{\bf{r}}_m}} \right)}}{{\partial {y_m}\partial {x_m}}}{\rm{ }}\frac{{{\partial ^2}f\left( {{{\bf{r}}_m}} \right)}}{{\partial {y_m}\partial {y_m}}}
	\end{array} \right]$ are detailed in (\ref{eqn73})-(\ref{eqn78}).

	\begin{figure*}[ht]
	\begin{footnotesize}
		\begin{align}
			\frac{{\partial f\left( {{{\bf{r}}_m}} \right)}}{{\partial {x_m}}} &= - \frac{{2\pi }}{\lambda }{\left| {{\alpha _d}} \right|^2}\sum\limits_{n = 1}^N {\sum\limits_{j = 1}^{L_d^r} {\sum\limits_{p = 1}^{L_d^r} {\left[ {{{\left| {{u_m}} \right|}^2}\left| {{{\left[ {{{\widetilde {\bf{G}}}_{d,n}}} \right]}_{j,p}}} \right|\sin  {\lambda _{d,n,j,p}^m}  \dot \theta _{j,p}^d - \!\!\!\!\!\! \sum\limits_{b = 1,b \ne m}^M \!\!\!\!\!\! {2\left| {{u_m}} \right|\left| {{u_b}} \right|\left| {{{\left[ {{{\widetilde {\bf{G}}}_{d,n}}} \right]}_{j,p}}} \right|\sin  {\lambda _{d,n,j,p}^b}  \sin \phi _{d,j}^r\cos \theta _{d,j}^r} } \right]} } }, \label{eqn69} \\
			\frac{{\partial f\left( {{{\bf{r}}_m}} \right)}}{{\partial {y_m}}} &= - \frac{{2\pi }}{\lambda }{\left| {{\alpha _d}} \right|^2}\sum\limits_{n = 1}^N {\sum\limits_{j = 1}^{L_d^r} {\sum\limits_{p = 1}^{L_d^r} {\left[ {{{\left| {{u_m}} \right|}^2}\left| {{{\left[ {{{\widetilde {\bf{G}}}_{d,n}}} \right]}_{j,p}}} \right|\sin {\lambda _{d,n,j,p}^m}  \ddot \theta _{j,p}^d - \!\!\!\!\!\! \sum\limits_{b = 1,b \ne m}^M \!\!\!\!\!\! {2\left| {{u_m}} \right|\left| {{u_b}} \right|\left| {{{\left[ {{{\widetilde {\bf{G}}}_{d,n}}} \right]}_{j,p}}} \right|\sin   {\lambda _{d,n,j,p}^b}  \cos \phi _{d,j}^r} } \right]} } }, \label{eqn70} \\
			\frac{{\partial g\left( {{{\bf{r}}_m}} \right)}}{{\partial {x_m}}} &= - \frac{{2\pi }}{\lambda }\sum\limits_{l = 1}^L  {\left| {{\alpha _l}} \right|^2}\sum\limits_{n = 1}^N {\sum\limits_{j = 1}^{L_l^r} {\sum\limits_{p = 1}^{L_l^r} {\left[ {{{\left| {{u_m}} \right|}^2}\left| {{{\left[ {{{\widetilde {\bf{G}}}_{l,n}}} \right]}_{j,p}}} \right|\sin  {\lambda _{l,n,j,p}^m}  \dot \theta _{j,p}^l - \!\!\!\!\!\! \sum\limits_{b = 1,b \ne m}^M \!\!\!\!\!\! {2\left| {{u_m}} \right|\left| {{u_b}} \right|\left| {{{\left[ {{{\widetilde {\bf{G}}}_{l,n}}} \right]}_{j,p}}} \right|\sin   {\lambda _{l,n,j,p}^b}  \sin \phi _{l,j}^r\cos \theta _{l,j}^r} } \right]} } }, \label{eqn71} \\
			\frac{{\partial g\left( {{{\bf{r}}_m}} \right)}}{{\partial {y_m}}} &= - \frac{{2\pi }}{\lambda }\sum\limits_{l = 1}^L {{{\left| {{\alpha _l}} \right|}^2}\sum\limits_{n = 1}^N {\sum\limits_{j = 1}^{L_l^r} {\sum\limits_{p = 1}^{L_l^r} {\left[ {{{\left| {{u_m}} \right|}^2}\left| {{{\left[ {{{\widetilde {\bf{G}}}_{l,n}}} \right]}_{j,p}}} \right|\sin   {\lambda _{l,n,j,p}^m}  \ddot \theta _{j,p}^l - \!\!\!\!\!\! \sum\limits_{b = 1,b \ne m}^M \!\!\!\!\!\! {2\left| {{u_m}} \right|\left| {{u_b}} \right|\left| {{{\left[ {{{\widetilde {\bf{G}}}_{l,n}}} \right]}_{j,p}}} \right|\sin   {\lambda _{l,n,j,p}^b}  \cos \phi _{l,j}^r} } \right]} } } }. \label{eqn72} \\
			\frac{{{\partial ^2}f\left( {{{\bf{r}}_m}} \right)}}{{\partial x_m^2}} &= - \frac{{4{\pi ^2}}}{{{\lambda ^2}}}{\left| {{\alpha _d}} \right|^2}\sum\limits_{n = 1}^N {\sum\limits_{j = 1}^{L_d^r} {\sum\limits_{p = 1}^{L_d^r}  \left[ {{{\left| {{u_m}} \right|}^2}\left| {{{\left[ {{{\widetilde {\bf{G}}}_{d,n}}} \right]}_{j,p}}} \right|\cos   {\lambda _{d,n,j,p}^m}  {{\left( {\dot \theta _{j,p}^d} \right)}^2} + \!\!\!\!\!\! \sum\limits_{b = 1,b \ne m}^M \!\!\!\!\!\! {2\left| {{u_m}} \right|\left| {{u_b}} \right|\left| {{{\left[ {{{\widetilde {\bf{G}}}_{d,n}}} \right]}_{j,p}}} \right|\cos   {\lambda _{d,n,j,p}^b}  {{\sin }^2}\phi _{d,j}^r{{\cos }^2}\theta _{d,j}^r} } \right]} }, \label{eqn73} \\
			\frac{{{\partial ^2}f\left( {{{\bf{r}}_m}} \right)}}{{\partial y_m^2}} &= - \frac{{4{\pi ^2}}}{{{\lambda ^2}}}{\left| {{\alpha _d}} \right|^2}\sum\limits_{n = 1}^N {\sum\limits_{j = 1}^{L_d^r} {\sum\limits_{p = 1}^{L_d^r} {\left[ {{{\left| {{u_m}} \right|}^2}\left| {{{\left[ {{{\widetilde {\bf{G}}}_{d,n}}} \right]}_{j,p}}} \right|\cos   {\lambda _{d,n,j,p}^m}  {{\left( {\ddot \theta _{j,p}^d} \right)}^2} \! + \!\!\!\!\!\! \sum\limits_{b = 1,b \ne m}^M \!\!\!\!\!\! {2\left| {{u_m}} \right|\left| {{u_b}} \right|\left| {{{\left[ {{{\widetilde {\bf{G}}}_{d,n}}} \right]}_{j,p}}} \right|\cos   {\lambda _{d,n,j,p}^b}  {{\cos }^2}\phi _{d,j}^r} } \right]} } }, \label{eqn74} \\
			\frac{{{\partial ^2}f\left( {{{\bf{r}}_m}} \right)}}{{\partial {x_m}\partial {y_m}}} \!\! &= \!  - \frac{{4{\pi ^2}}}{{{\lambda ^2}}}\!\!{\left| {{\alpha _d}} \right|^2} \!\! \sum\limits_{n = 1}^N \!\!{\sum\limits_{j = 1}^{L_d^r} \!\! {\sum\limits_{p = 1}^{L_d^r}\!\! {\left[ \! {{{\left| {{u_m}} \right|}^2}\left|\! {{{\left[\! {{{\widetilde {\bf{G}}}_{d,n}}} \right]}_{j,p}}} \right|\cos  {\lambda _{d,n,j,p}^m}  \dot \theta _{j,p}^d\ddot \theta _{j,p}^d \! + \!\!\!\!\!\!\! \sum\limits_{b = 1,b \ne m}^M \!\!\!\!\!\! {2\left| {{u_m}} \right|\left| {{u_b}} \right|\left| {{{\left[ {{{\widetilde {\bf{G}}}_{d,n}}} \right]}_{j,p}}} \right|\cos   {\lambda _{d,n,j,p}^b}  \sin \phi _{d,j}^r\cos \theta _{d,j}^r\cos \phi _{d,j}^r} } \right]} } } , \label{eqn75} \\
			\frac{{{\partial ^2}g\left( {{{\bf{r}}_m}} \right)}}{{\partial x_m^2}} \! &= \!  - \frac{{4{\pi ^2}}}{{{\lambda ^2}}}\sum\limits_{l = 1}^L  {\left| {{\alpha _l}} \right|^2}\sum\limits_{n = 1}^N {\sum\limits_{j = 1}^{L_l^r}  \sum\limits_{p = 1}^{L_l^r}  \left[ {{{\left| {{u_m}} \right|}^2}\left| {{{\left[ {{{\widetilde {\bf{G}}}_{l,n}}} \right]}_{j,p}}} \right|\cos   {\lambda _{l,n,j,p}^m}  {{\left( {\dot \theta _{j,p}^l} \right)}^2} \!+ \!\!\!\!\!\! \sum\limits_{b = 1,b \ne m}^M \!\!\!\!\!\!{2\left| {{u_m}} \right|\left| {{u_b}} \right|\left| {{{\left[ {{{\widetilde {\bf{G}}}_{l,n}}} \right]}_{j,p}}} \right|\cos  {\lambda _{l,n,j,p}^b}  {{\sin }^2}\phi _{l,j}^r{{\cos }^2}\theta _{l,j}^r} } \right]}, \label{eqn76} \\
			\frac{{{\partial ^2}g\left( {{{\bf{r}}_m}} \right)}}{{\partial y_m^2}} \! &= \! - \frac{{4{\pi ^2}}}{{{\lambda ^2}}}\sum\limits_{l = 1}^L {{{\left| {{\alpha _l}} \right|}^2}\sum\limits_{n = 1}^N {\sum\limits_{j = 1}^{L_l^r} {\sum\limits_{p = 1}^{L_l^r} {\left[ {{{\left| {{u_m}} \right|}^2}\left| {{{\left[ {{{\widetilde {\bf{G}}}_{l,n}}} \right]}_{j,p}}} \right|\cos  {\lambda _{l,n,j,p}^m}  {{\left( {\ddot \theta _{j,p}^l} \right)}^2} \! + \!\!\!\!\!\! \sum\limits_{b = 1,b \ne m}^M \!\!\! {2\left| {{u_m}} \right|\left| {{u_b}} \right|\left| {{{\left[ {{{\widetilde {\bf{G}}}_{l,n}}} \right]}_{j,p}}} \right|\cos  {\lambda _{l,n,j,p}^b}  {{\cos }^2}\phi _{l,j}^r} } \right]} } } }, \label{eqn77} \\
			\frac{{{\partial ^2}g\left( {{{\bf{r}}_m}} \right)}}{{\partial {x_m}\partial {y_m}}} \!\! &= \!  -  \frac{{4{\pi ^2}}}{{{\lambda ^2}}}\!\!\sum\limits_{l = 1}^L {{{\!\left| {{\alpha _l}} \right|}^2}\!\!\sum\limits_{n = 1}^N \! {\sum\limits_{j = 1}^{L_l^r}\! {\sum\limits_{p = 1}^{L_l^r} \!\!{\left[\!  {{{\left| {{u_m}} \right|}^2} \!\left| \!{{{\left[ \! {{{\widetilde {\bf{G}}}_{l,n}}} \right]}_{j,p}}} \right|\!\cos {\lambda _{l,n,j,p}^m}   \!\!\left(\! {\dot \theta _{j,p}^l\ddot \theta _{j,p}^l} \!\right) \!\!+\!\!\!\!\!\!\! \sum\limits_{b = 1,b \ne m}^M \!\!\!\!\!\!{2\!\left| {{u_m}} \right| \!\left| {{u_b}} \right| \!\left|\! {{{\left[ {{{\widetilde {\bf{G}}}_{l,n}}} \right]}_{j,p}}} \right| \! \cos   {\lambda _{l,n,j,p}^b}  \sin \phi _{l,j}^r\cos \theta _{l,j}^r\cos \phi _{l,j}^r} }\! \right]} } } }. \label{eqn78}
		\end{align}	 
	\end{footnotesize}

    	\hrulefill
    \end{figure*}

\end{appendices}



%

\end{document}